\documentclass[12pt,english]{article}
\pdfoutput=1
\usepackage{jheppub}
\usepackage{simplewick}
\usepackage{verbatim}
\usepackage{graphics, color}
\usepackage{graphicx}
\usepackage{amsmath}
\usepackage{amssymb}
\usepackage{amsthm}
\usepackage{amsfonts}
\usepackage[usenames,dvipsnames]{xcolor}
\usepackage[normalem]{ulem}
\usepackage{braket}

\newcommand{\be}{\begin{equation}}
\newcommand{\ee}{\end{equation}}

\newcommand{\lan}{\langle}
\newcommand{\ran}{\rangle}

\newcommand{\mO}{\mathcal{O}}

\definecolor{grey}{rgb}{.5,.5,.5}
\definecolor{bluegreen}{rgb}{0,.5,.5}
\definecolor{darkgreen}{rgb}{0,.5,0}

\newcommand{\beq}{\begin{equation}}
\newcommand{\eeq}{\end{equation}}
\def\({\left(}
\def\){\right)}

\begin{document}

\title{Escaping the Interiors of Pure Boundary-State Black Holes}
\author{Ahmed Almheiri$^{1,2}$}
\author{Alexandros Mousatov$^1$,}
\author{Milind Shyani$^1$}
\affiliation{$^1$Stanford Institute for Theoretical Physics, Department of Physics, Stanford University, Stanford, CA 94305, USA}
\affiliation{$^2$Institute for Advanced Study,  Princeton, NJ 08540, USA}
\emailAdd{almheiri@ias.edu}
\emailAdd{mousatov@stanford.edu}
\emailAdd{shyani@stanford.edu}
\abstract{We consider a  class of pure black hole microstates and demonstrate that they can be made escapable by turning on certain double trace deformations in the CFT. These microstates are dual to BCFT states prepared via a Euclidean path integral starting from a boundary in Euclidean time. These states are dual to black holes in the bulk with an End-of-the-World brane; a codimension one timelike boundary of the spacetime behind the horizon. We show that by tuning the sign of the coupling of the double trace operator to the boundary conditions on the brane the deformation injects negative energy into the black hole causing a time advance for signals behind the horizon. We demonstrate how the property of escapability in the considered microstates follows immediately from the traversability of deformed wormholes. We briefly comment on reconstruction of the black hole interior and state dependence.

}

\maketitle
\addtocontents{toc}{\protect\enlargethispage{2\baselineskip}}

\section{Introduction}

There is no more an abstruse a place than the interiors of pure black holes. Concealed behind the event horizon, the very existence of the interior has been put into question by thought experiments \cite{Almheiri:2012rt} motivated by the black hole information paradox \cite{Hawking:1976ra}. Several proposals have been put forth in an attempt to save the interior \cite{Papadodimas:2012aq, Papadodimas:2013jku, Papadodimas:2015jra, Verlinde:2012cy, Verlinde:2013uja, Verlinde:2013qya, Maldacena:2013xja} all of which cannot be confirmed nor invalidated explicitly\footnote{Except for the infalling observer.}. 

This question about the exitence of the interior becomes sharpest in the context of the AdS/CFT correspondence. It is sometimes reformulated as to whether we can define CFT operators dual to low energy bulk operators that probe the interiors of black holes. The program of bulk reconstruction is very well understood via the HKLL construction of local bulk operators as smeared boundary operators \cite{Hamilton:2006az}. The direct HKLL method has restricted use due to its dependence on causality, where the bulk operator to be reconstructed has to be causally connected to the boundary, i.e. both receive and send causal signals to the boundary. This limitation was recently superseded, in certain cases, with modified HKLL-like constructions aided by knowing the modular Hamiltonian of the state \cite{Faulkner:2017vdd} or using knowledge of how the state was constructed \cite{Almheiri:2017fbd}.

A major game changer in this story is the recent discovery that wormholes can be made traversable \cite{Gao:2016bin}. Starting with the thermofield double state dual to the eternal black hole, it was found that turning on a simple Hamiltonian double trace deformation of the two CFTs coupling a local operator on one CFT to one on the other can take advantage of the carefully tuned correlations in the thermofield double between the CFTs to insert negative energy into the bulk. This negative energy induces a violation of the average null energy condition (ANEC) on the horizon, and allows for signals to escape the interior of the wormhole. A more direct probe of this traversability was studied in \cite{Maldacena:2017axo} which showed how the deformation implies a non-vanishing commutator between operators in the two CFTs, consistent with a signal passing through the wormhole. These results imply that points in the interior of the wormhole are now causally connected to the boundary, potentially allowing for a new HKLL-like prescription to find CFT representations of interior bulk operators\footnote{Although, one might argue that we could have past-evolved the fields from inside the black hole to initial data on a time slice intersecting the bifurcate horizon and then used the HKLL construction. The traversability doesn't really buy us anything new.}.

In this work we present an analogous effect in the case of a single sided black hole; We find a hamiltonian deformation which makes a class of pure black holes escapable. This work is inspired by the recent paper \cite{Kourkoulou:2017zaj} which demonstrated this effect for pure black hole microstates in 1+1 dilaton gravity dual to the SYK model. In higher dimensions, we consider a special class of black hole microstates where the state is specified by properties of an end-of-the-world brane located behind the event horizon. We focus only on the small class of states specified by the boundary conditions of the bulk fields on the brane. These states are dual to Cardy boundary states; Pure CFT states with a boundary in Euclidean time preserving (some) conformal symmetry. We show how a simple Hamiltonian deformation, composed of the integral of a local double trace operator, injects negative energy through the horizon making it escapable. We find that the deformation needs to be `state-dependent', where the sign of the coupling needs to be tuned accordingly with the boundary conditions on the brane.

\section{Boundary State Black Holes}

In this section, we describe the construction of the black hole microstates of interest. We present their construction from the boundary perspective involving a conformal field theory with a boundary in Euclidean time, and from the bulk side as a black hole spacetime where the boundary's boundary extends into the bulk as an end of the world D-brane. The general dimensional case will be considered in describing the set up and then we'll focus on $2+1$ dimensional bulk for explicit calculations.

\subsection{Boundary States}

This section is based on reviews \cite{Cardy:2004hm, Liendo:2012hy, McAvity:1995zd}.

Consider a Euclidean CFT in $d$ dimensions on a manifold with a $d-1$ dimensional boundary that perserves some conformal symmetry, i.e. a BCFT. Let the topology of the space be $R^d$ and of the boundary to be $R^{d-1}$. Since the location of boundary needs to be preserved under conformal transformations, some of the symmetries are broken and the original $SO(d,1)$ conformal symmetry reduces to $SO(d-1,1)$\footnote{Also, since the boundary is $\mathbb{R}^{d-1}$ it is by itself $SO(d-1,1)$ invariant.}.

The simplest case to consider is a BCFT on a manifold whose metric is
\begin{align}
ds^2 = d\tau^2 + d\vec{x}^2,
\end{align}
and with domain $\tau \ge 0$ (we will call this domain the `\emph{Upper Half Plane}', UHP). It can be shown, using the conformal ward identities in the presence of the boundary $\tau = 0$, that the stress tensor has to satisfy $T_{\tau x^i} (0,\vec{x}) = 0$ at the boundary. This can be thought of as the requirement that  no `energy/momentum' is lost across the boundary. 

Correlation functions are also quite interesting in the presence of the boundary. Take for instance the one point function of a local operator. In the case of a CFT on the plane the one point function of a primary operator in the vacuum is required to vanish by translation invariance. This symmetry is broken in a BCFT. Due to the remaining conformal symmetry the one point function attains the form
\begin{align}
\lan \mO(\tau,\vec{x}) \ran_{UHP} = {A_\mO \over (2 \tau)^\Delta},
\end{align}
where $A_\mO$ is determined by the details of the theory and the precise boundary state in question. One could think of this as the boundary providing a source for the operator $\mO$; we review the bulk interpretation of this below. 

The two point function of a primary operator in a BCFT is more complicated than the case with no boundaries (where it's exactly fixed by the symmetries). Non-trivial information about the operator content and  OPE coefficients is necessary to compute the two point function exactly in a BCFT. Since the one point function of primaries no longer vanish, the two point function receives contributions from primaries other than just the identity operator. Thus, 
\begin{align}
\lan \mO(x_1) \mO(x_2) \ran_{UHP} \sim \sum_{p} C_{\mO \mO}^{\mO_p} F(x_1, x_2, \partial_1) \lan \mO_p(x_1) \ran_{UHP}
\end{align}
However, working with a holographic theory which admits large $N$ factorization simplifies the problem considerably. At large $N$, the sum above is dominated by the identity operator and double trace operators of the form $\mO \partial^k \mO$. The final result is given by a slightly modified version of Wick's theorem\cite{Liendo:2012hy}:
\begin{align} 
\lan \mO(x_1) \mO(x_2) \ran_{UHP} = \lan \mO(x_1) \ran_{UHP} \lan \mO(x_2) \ran_{UHP} + \lan \mO(x_1) \mO(x_2) \ran  \pm \lan \mO(x_1) \mO(x_2^* ) \ran \label{wick}
\end{align}
where
\begin{align}
\lan \mO(x_1) \mO(x_2) \ran  = {1 \over |x_1 - x_2|^{2\Delta}}
\end{align}
and by $x^*$ we mean that the sign of $\tau$ component is flipped. The intuitive reason for this is that the boundary behaves like a mirror and the operator with the flipped $\tau$ coordinate plays the role of the mirror charge. The sign of the last term in eq. \ref{wick} is governed by the boundary condition of $\mO$ at the boundary, being either Dirchlet ($-$) or  Neumann ($+$). In the AdS bulk, this information is controlled by the black hole microstate.

The boundary state considered here is somewhat close to the vacuum, as can be seen by the form of the correlation functions. Boundary states of higher energy can be obtained easily in $d = 2$ by a simple conformal transformation. By first working in the complex coordinate system as $z = i \tau + x, \bar{z} = - i \tau + x$ we consider the coordinate transformation
\begin{align}
z \rightarrow w = -i  {\beta / 4} + {\beta \over 2 \pi} \ln z. \label{coordtransf}
\end{align}
This transformation maps the UHP to a strip of width $\beta/2$, mapping the positive (negative) real axis to the lower (upper) edge of the strip. Correlation functions in this new domain can now be thought of as expectation values in the state obtained by evolving a boundary state by $\beta/4$ euclidean time,
\begin{align}
| B_\beta \rangle = e^{- {\beta \over 4} H} | B \ran
\end{align}
\begin{figure}
\begin{center}
\includegraphics[height=4cm]{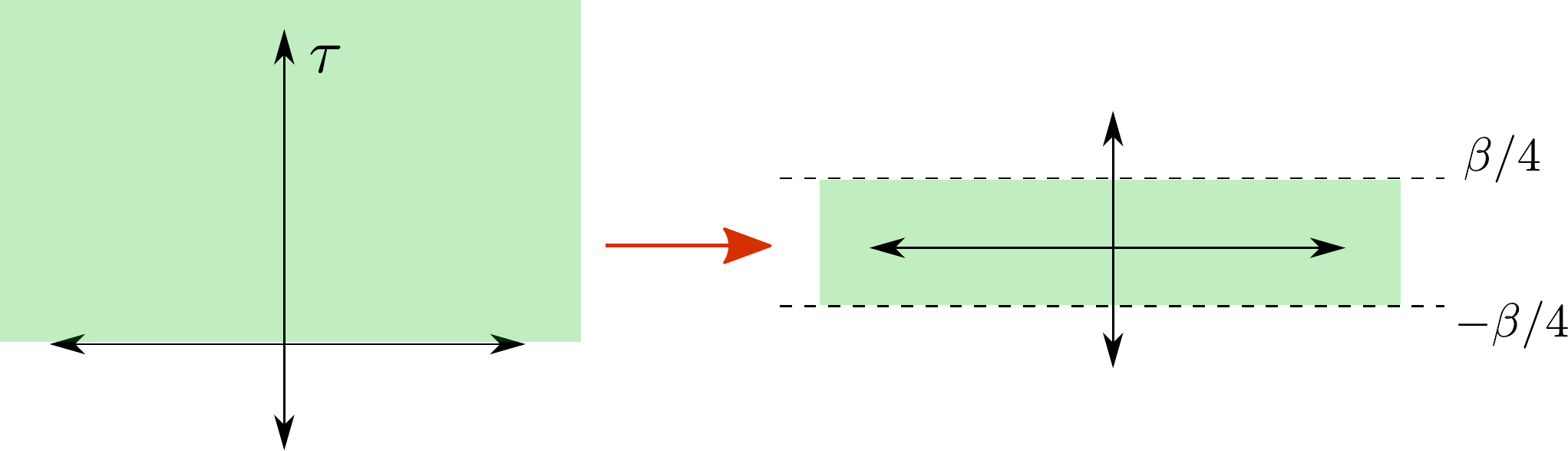}
\caption{The coordinate transformation \ref{coordtransf} maps the UHP to the strip. The state $| B \ran$ should be thought as prepared on the boundaries of the strip.}\label{UHPtoStrip}
\end{center}
\end{figure}Since primary operators continue to transform in the usual way, the correlation functions  \ref{wick} now transform to
\begin{align}
\lan \mO(w) \ran_{strip} &= {A_\mO \over \left( {\beta \over \pi} \cos\left[  {2 \pi \over \beta} \mathrm{Im}[w]  \right]    \right)^\Delta}  \label{btzcorrelators1}\\
\lan \mO(w_1) \mO(w_2) \ran_{strip}^{connected} &= {1 \over \Big| {\beta \over \pi} \sinh\left[ {\pi \over \beta} (w_1 - w_2) \right] \Big|^{2 \Delta}} \pm {1 \over \Big| {\beta \over \pi} \cosh\left[ {\pi \over \beta} (w_1 - \bar{w}_2) \right] \Big|^{2 \Delta}} \label{btzcorrelators2}
\end{align}
where the second line is only the connected piece of the large $N$ two point function. Note how taking $\beta \rightarrow \infty$ we obtain the UHP BCFT result.

\subsection{Bulk Dual of Boundary States}

In this paper we assume the proposal of \cite{Takayanagi:2011zk, Fujita:2011fp} in the construction of the holographic dual of a boundary CFT, of which this section is mostly a review. Other related work include \cite{DeWolfe:2001pq, Alishahiha:2011rg}. 

The general idea is to consider an AdS space with two boundaries, one being the asymptotic holographic boundary and the other being the bulk extension of the boundary in the CFT. The latter boundary is to be thought of as an end-of-the-world (ETW) D-brane emanating from the asymptotic boundary into the bulk. The action governing this system is taken to be
\begin{align}
S = {1 \over 16 \pi G} \int\displaylimits_{bulk}\! \! \!  d^{d+1}x \sqrt{g} \left( R - 2 \Lambda \right) + {1 \over 8 \pi G} \! \! \! \! \! \!  \int\displaylimits_{boundary} \quad \! \! \! \! d^{d}x \sqrt{h} K + {1 \over 8 \pi G} \! \! \! \int\displaylimits_{brane} \! \! \!  d^{d}x \sqrt{h} \left( K - T \right) \label{action}
\end{align}
where $K$ is the extrinsic curvature and $T$ is the brane tension\footnote{Other possible terms include dynamical fields on the brane that may arise from string theory.}. Note how there are two extrinsic curvature terms one for each boundary: the brane and the asymptotic boundary.

We impose Dirichlet boundary conditions on the metric at the asymptotic boundary, but impose Neumann conditions on the brane. This allows the brane to be dynamical in the AdS bulk. A well defined variational principle for the action \ref{action} near the brane requires the  Neumann boundary condition on the metric \cite{Takayanagi:2011zk, Fujita:2011fp} 
\begin{align}
K_{ab} = (K-T) h_{ab}, \label{ext}
\end{align}
the trace of which gives,
\begin{align}
K = {d \over d - 1} T.
\end{align} 
This constraint on the extrinsic curvature along with a solution of the bulk equations of motion determines the trajectory of the brane on this background.

\begin{figure}
\begin{center}
\includegraphics[height=4cm]{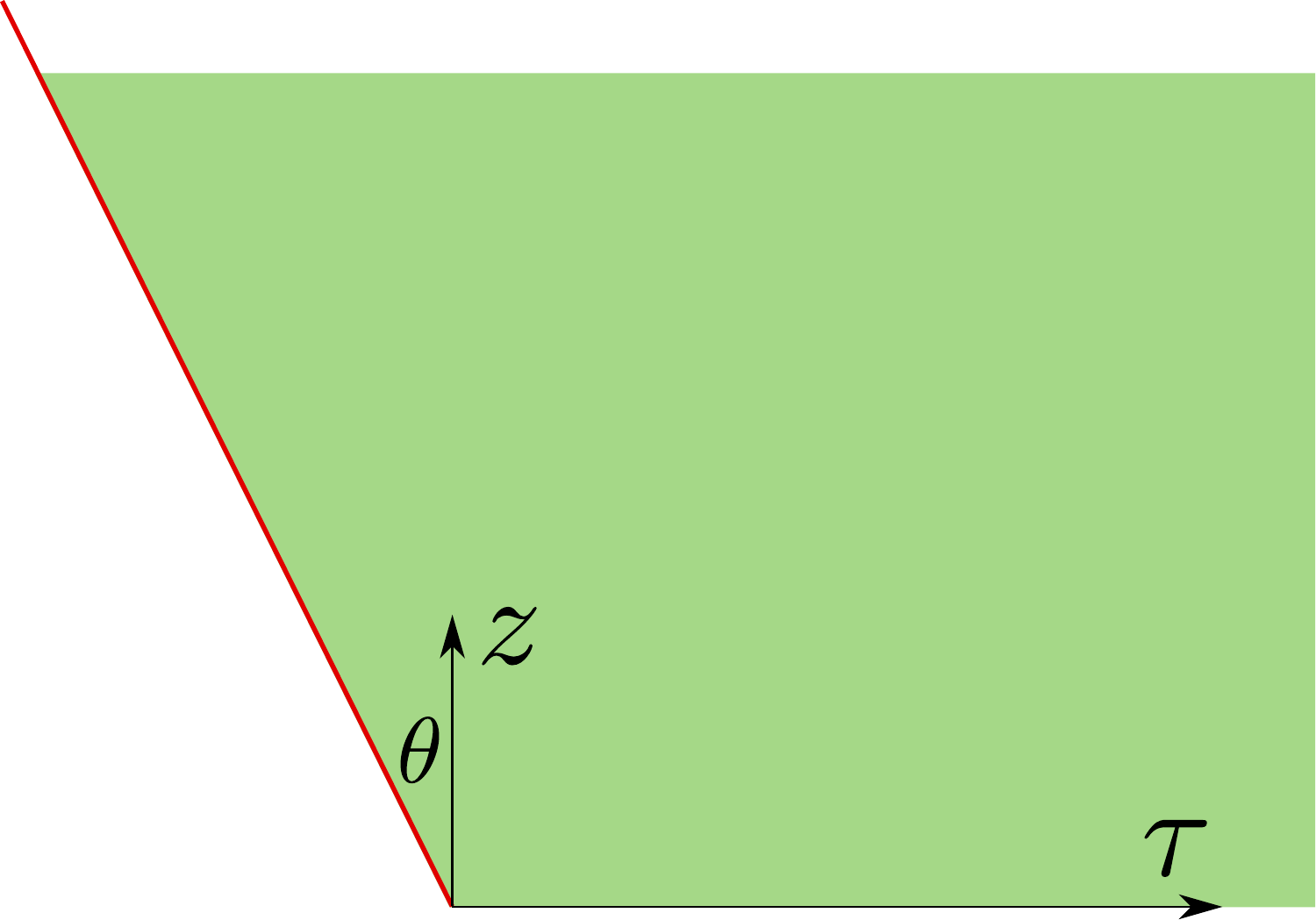}
\caption{The bulk (green) is bounded by the asymptotic boundary at $z = 0$, and the brane (red) shooting off at an angle $\theta$ in the $z-\tau$ plane, hits the boundary at $\tau=0$. Transverse directions are suppressed.}\label{PoincareBrane}
\end{center}
\end{figure}

Let's consider a few examples. Starting with a bulk given by Euclidean Poincare AdS, different brane tensions $T$ correspond to different angles in the $t-z$ plane, see figure \ref{PoincareBrane}. Vanishing tension corresponds to $\theta = 0$.

We are interested in black hole spacetimes with an ETW brane. We therefore look for solutions of \ref{ext} in the background of a Euclidean black hole. The behavior is qualitatively similar for all dimensions, but we focus on Euclidean BTZ. We find that the brane is always anchored at antipodal points of the thermal circle, and its trajectory, $r(\tau)$, satisfies
\begin{align}
{T \over \sqrt{1 - T^2}} = - \cos(r_+ \tau) {\sqrt{r^2(\tau) - r_+^2} \over r_+}, \label{EuclideanBrane}
\end{align}
in the coordinates of the BTZ metric
\begin{align}
ds^2 = (r^2 - r_+^2) d\tau^2 + {dr^2 \over r^2 - r_+^2} + r^2 dx^2.
\end{align}
where $\beta = 2 \pi/ r_+$. Continuing to Lorenztian time, and reorganizing \ref{EuclideanBrane}, gives the trajectory
\begin{align}
r(t) = {r_+ \over \sqrt{1 - T^2}} \sqrt{1 - T^2 \tanh^2[2 \pi t/\beta]} \label{LorentzianBrane}
\end{align}
Which side of the ETW brane correponds to the `interior world' is determined by the sign of the tension of the brane; the sign of the tension determines whether we pick a given normal vector or its opposite.

\begin{figure}
\begin{center}
\includegraphics[height=4.5cm]{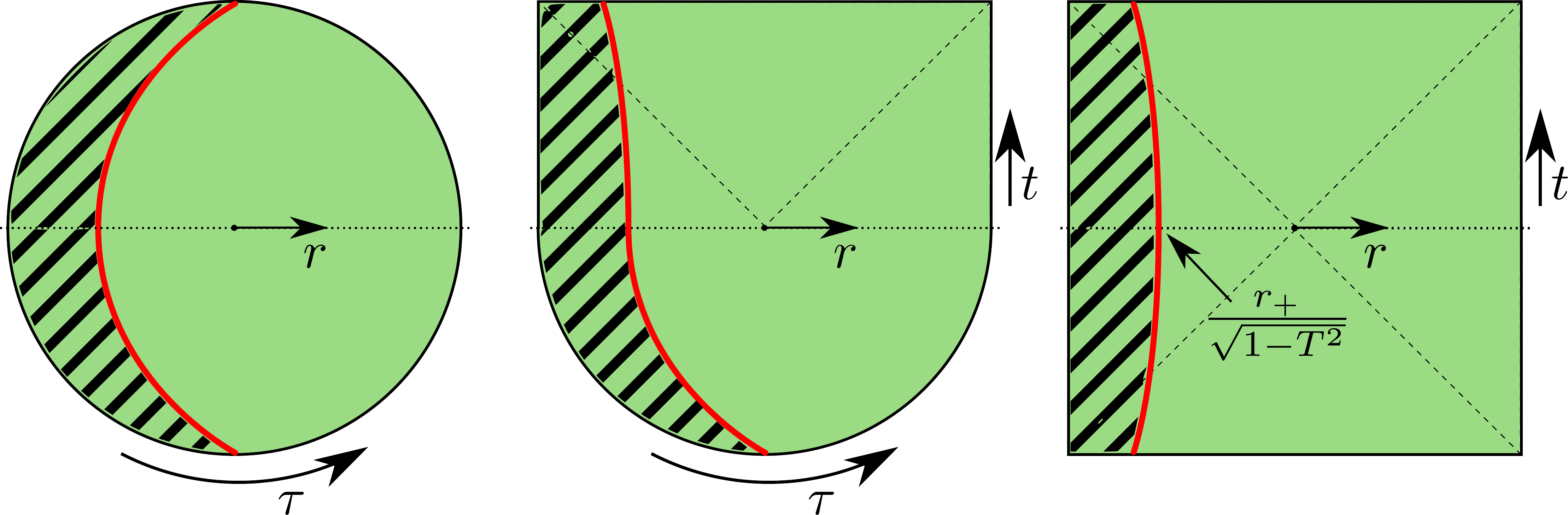}
\caption{Allowed positive tension brane configurations in the Euclidean/Lorenztian BTZ geometry. Left: The trajectory of the End-of-The-World (ETW) brane (red) in the Euclidean geometry. Middle: Euclidean preparation of the Lorentzian ETW brane BTZ state. Right: Full Lorenztian solution of ETW brane BTZ black hole. The shaded region is the excluded region of the original eternal BTZ black hole solution.}\label{BHBrane}
\end{center}
\end{figure}

\begin{figure}
\begin{center}
\includegraphics[height=4cm]{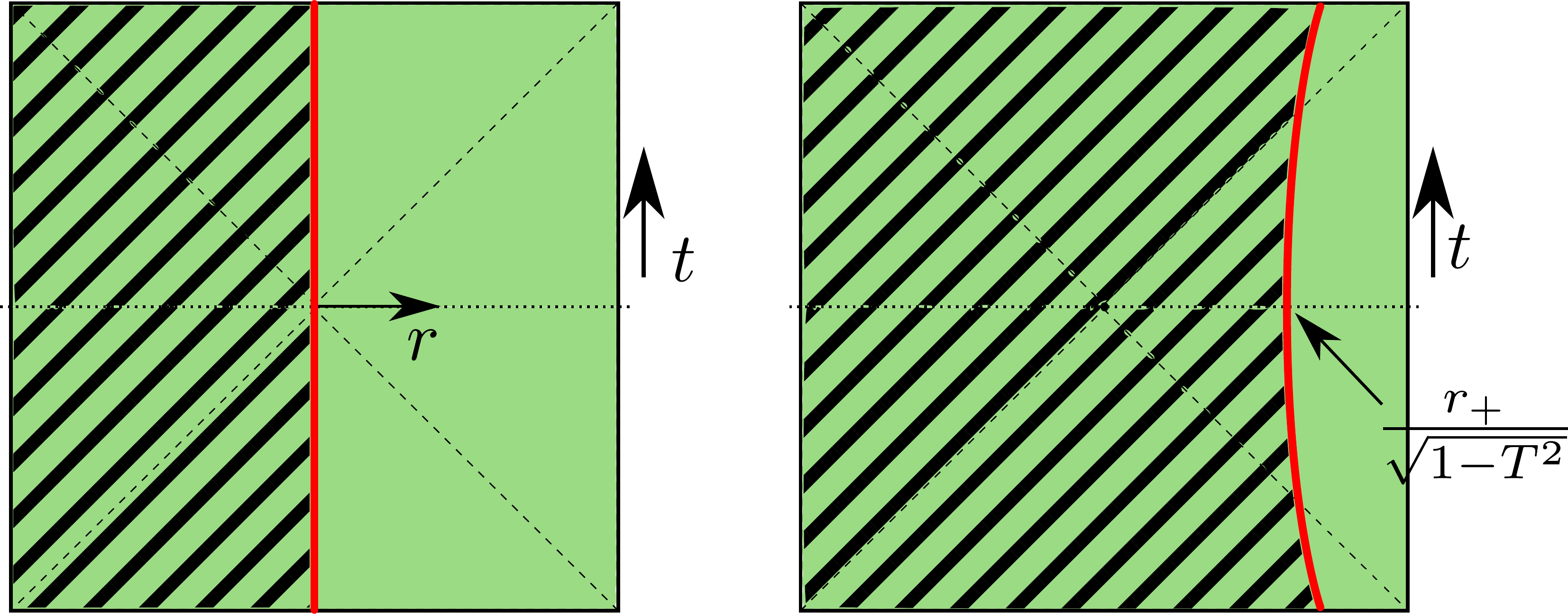}
\caption{Left: Zero tension ETW brane. The solution can be thought of as a $Z_2$ orbifold of the original eternal black hole solution. Right: Negative tension ETW brane. From the right exterior this solution looks like a brane which emerges from the white hole and falls into the black hole.}\label{BHBraneZN}
\end{center}
\end{figure}

In Euclidean BTZ we see from \ref{EuclideanBrane} that the brane is anchored on the boundary at $\tau = 0, \beta/2$, and having non-zero tension simply displaces the brane away from the tip of the Euclidean cigar at $r = r_+$. The closest distance between the brane and the tip is $r_+/\sqrt{1 - T^2}$. These solutions are valid only for $0 \le |T| <1$. The lorentzian space solution involves a brane which emerges out of the white hole singularity and crashes into the black hole singularity for all allowed values of $T$. For positive tension, this state describes a single sided black hole with ``lots of interior'' behind the original bifurcation surface of the black hole. In contrast, the negative Tension solutions completely excise the bifurcation horizon and describes a brane emerging from the white hole which then falls back into the black hole. The zero tension solution simply cuts out half of the eternal black hole spacetime. These solutions are shown in figures \ref{BHBrane} and \ref{BHBraneZN} .

\subsubsection{Brane Microstates}

Branes with different tension in the bulk correspond to different boundary states in the CFT. We will restrict our attention to the case of zero tension, where the brane shoots off at a right angle from the boundary. We will discuss here the relation between the brane microstates and the boundary conditions of the bulk fields on the brane arises.

To analyze this, consider the free scalar action in Euclidean Poincare $AdS_{d+1}$ with an ETW brane boundary:
\begin{align}
S = {1 \over 2} \int d^{d+1} x \sqrt{g}  \left[ (\nabla \Phi)^2 + m^2 \Phi^2 \right]
\end{align}
with metric 
\begin{align}
ds^2 = {dz^2 + d\tau^2 + d\vec{x}^2\over z^2}
\end{align}
with domain $\tau,z \ge 0$ and the brane is located at $\tau = 0$. The presence of the brane introduces new contributions to the above free action, coming from the Dirac-Borne-Infled action, which govern how the open strings on the brane interact with the closed strings in the bulk. This includes terms of the form
\begin{align}
S_{brane} = \int d^{d}x \sqrt{h} A \Phi + ...
\end{align}
where $h$ is the pullback of the metric on the brane and $A$ is a constant coming from VEVs of the fields living on the brane; it is some background excitation of the open strings on the brane. This term provides a source for the scalar field in the bulk.

Assuming these to be the only relevant terms at leading order in $N$, one can consider the variation of the scalar action, giving
\begin{align}
\delta S + \delta S_{brane} = \int dz d^{d-1}\vec{x} {1 \over z^{d-1}} \delta \Phi \left( \partial_\tau \Phi + {A \over z}\right) + \int d\tau d^{d-1}\vec{x} {1 \over z^{d-1}} \delta \Phi \partial_z \Phi.
\end{align}
where the first term corresponds to the brane boundary, and the second term to the asymptotic boundary. We therefore need to pick two boundary conditions. The second term involves the standard choices in the holographic dictionary, for which we choose the usual Dirichlet condition $\delta \phi = 0$ (or possibly $\partial_z \phi = 0$ for the alternate quantization scheme). As for the first term, we have the choice between Dirichlet or a slightly modified Neumann condition
\begin{align}
\partial_\tau \Phi + {A \over z} = 0.
\end{align}
This gives a naive counting of the number of brane microstates in this sector of vanishing tension to be $2^K$ where $K$ is the number of bulk fields; two choices, Dirichlet or Neumann, for each bulk field. However, these states are most probably not orthogonal and computing their overlap goes beyond the tools developed in this paper. This subtlety does not play any major role in the present discussion.

\subsubsection{Boundary State Holography}

Holography in the presence of a boundary proceeds in the usual way while keeping track of the presence of a background value for the scalar. We focus on the Euclidean Poincare case with a zero tension brane. We write the bulk scalar field as
\begin{align}
\Phi = \phi_B + \phi
\end{align}
both terms satisfy the same wave equation separately. The general solution is
\begin{align}
\Phi(z,\tau,\vec{x}) = z^{d/2} \int dw d^{d-1}\vec{k} C_{w k} K_\nu(q z) e^{i w \tau} e^{i \vec{k} \cdot \vec{x}} 
\end{align}
where $\nu = \Delta - {d \over 2}$, $q^2 = w^2 + \vec{k}^2$, and $C_{w k}$ are the mode coefficients. We write here that $C_{wk} = c^B_{wk} + c_{wk}$ the mode functions of $\phi_B$ and $\phi$ respectively.

Let's consider first the Dirichlet boundary condition, say $\phi_B, \phi = 0$ at $\tau = 0$. This restricts the mode functions to be odd in $w$, and we see that  the background field $\phi_B$ can be consistently set to zero. The general solution in position space for the perturbation can be written in terms of a boundary source as
\begin{align}
\phi(z,\tau, \vec{x}) = z^{d/2} \int d\tau' d^{d-1}\vec{x}' \phi_0(\tau', \vec{x}') {\cal K}_D(z,\tau,\vec{x}; \tau', \vec{x}')
\end{align}
where 
\begin{align}
{\cal K}_D(z,\tau,\vec{x}; \tau', \vec{x}') = K(z,\tau,\vec{x}; \tau', \vec{x}')  - K(z,\tau,\vec{x}; -\tau', \vec{x}') 
\end{align}
where $K$ is the usual pure $AdS$ bulk to boundary propagator given by
\begin{align}
K(z,\tau,\vec{x}; \tau', \vec{x}') &= z^{d/2}  \int dw d^{d-1}\vec{k}  K_\nu(q z) e^{i w (\tau - \tau')} e^{i \vec{k} \cdot (\vec{x} - \vec{x}')}  \\
&= C {z^{\Delta} \over (z^2  + (\tau - \tau')^2+ (\vec{x} - \vec{x}')^2)^{ \Delta}},
\end{align}
and the source is
\begin{align}
\phi_0(\tau', \vec{x}')  = \int dw d^{d - 1}\vec{k} \ c_{w k} e^{ i w \tau'} e^{ i \vec{k} \cdot \vec{x}'}.
\end{align}
The Neumann case is a bit more interesting. Requiring that $\phi_B$ satisfy the modified Neumann equation $\partial_\tau \phi_B = -A/z$, completely determines the coefficients to be
\begin{align}
c^B_{w k} =  -\delta^{d-1}(k) {|w|^{d/2} \over w} \left( \frac{A}{2^{\frac d 2 - 1}\Gamma\left(\frac{1+\frac d 2+\nu}{2}\right)\Gamma\left(\frac{1+\frac d 2-\nu}{2}\right)} \right).
\end{align}
In fact, this completely fixes the background field value to be
\begin{align}
\phi_B \propto \left( {\tau \over z}\right) {}_2F_1\left({\Delta + 1 \over 2}, {d - \Delta + 1 \over 2}, {3 \over 2}; - {\tau^2 \over z^2}\right)
\end{align}
which upon expanding near $z\sim 0$ gives
\begin{align}
\phi_B \sim C_1 \left( z \over \tau \right)^{d-\Delta} (1 + ...) + C_2 \left( {z \over \tau} \right)^\Delta (1 + ....)\end{align}
where $C_{1,2}$ are known functions of $A$. In order for the Neumann condition to be completely satisfied, the perturbation needs to satisfy the simpler condition
\begin{align}
\partial_\tau \phi = 0
\end{align}
which forces the mode functions $c_{w k}$ to be even in $w$. Therefore, the solution in terms of a boundary source will be
\begin{align}
\phi(z,\tau, \vec{x}) = z^{d/2} \int d\tau' d^{d-1}\vec{x}' \phi_0(\tau', \vec{x}') {\cal K}_N(z,\tau,\vec{x}; \tau', \vec{x}')
\end{align}
where 
\begin{align}
{\cal K}_N(z,\tau,\vec{x}; \tau', \vec{x}') = K(z,\tau,\vec{x}; \tau', \vec{x}')  + K(z,\tau,\vec{x}; -\tau', \vec{x}') .
\end{align}

Having the form of the bulk to boundary propagator we can compute the generating functional for $\mO$, the operator dual to the normalizable mode of $\phi$. Starting with the scalar action we write
\begin{align}
S &= {1 \over 2} \int d\tau d^{d-1}\vec{x} dz \left(  {1 \over z^{d-1}} \left( \partial \phi_B + \partial {\phi}  \right)^2 + {m^2\over z^{d+1}} \left( \phi_B + \phi  \right)^2 \right) + S_{brane}\\
&= {1 \over 2} \int d\tau d^{d-1}\vec{x} dz \left( {1 \over z^{d-1}} \left( \partial \phi_B\partial \phi_B +   2 \partial {\phi} \partial \phi_B +  \partial {\phi}  \partial {\phi}    \right) + {m^2\over z^{d+1}} \left( \phi_B + \phi  \right)^2 \right) + S_{brane}
\end{align}
We disregard terms quadratic in $\phi_B$ since they wont contribute to the correlation functions of $\mO$. Integrating the rest by parts and using the equations of motion, we are left with
\begin{align}
S = {1 \over 2} \int d\tau d^{d-1}\vec{x} {1 \over z^{d-1}}\left(  2  \phi \partial_z \phi_B +       \phi \partial_z \phi            \right).
\end{align}
Note that the boundary terms on the brane vanish by the boundary conditions discussed earlier. Near the boundary we have
\begin{align}
\phi(z,\tau,\vec{x}) = z^{d - \Delta} \phi_0(\tau, \vec{x}) + z^{\Delta} \int d\tau' d^{d-1}\vec{x}' \phi_0(\tau', \vec{x}') \lan \mO(\tau, \vec{x}) \mO(\tau', \vec{x}') \ran_S
\end{align}
where $S \in \{D,N\}$ denotes whether the Dirichlet or Neumann condition has been imposed, and
\begin{align}
\lan \mO(\tau, \vec{x}) \mO(\tau', \vec{x}') \ran_S &= {1 \over \left( (\tau - \tau')^2 + (\vec{x} - \vec{x}')^2 \right)^\Delta}  - {1 \over \left( (\tau + \tau')^2 + (\vec{x} - \vec{x}')^2 \right)^\Delta},  \ \ S = D \\
&= {1 \over \left( (\tau - \tau')^2 + (\vec{x} - \vec{x}')^2 \right)^\Delta}  + {1 \over \left( (\tau + \tau')^2 + (\vec{x} - \vec{x}')^2 \right)^\Delta},  \ \ S = N
\end{align}
After the usual process of holographic renormalization we end up with
\begin{align}
S_{ren} = \!  \! \int \! d\tau d^{d-1}\vec{x}      \phi_0(\tau, \vec{x}) {C_2 \Delta \over \tau^\Delta} +    {1 \over 2} \! \! \int \! \! d\tau d^{d-1}\vec{x}    \phi_0(\tau, \vec{x}) \phi_0(\tau', \vec{x}')  \lan \mO(\tau, \vec{x}) \mO(\tau', \vec{x}') \ran_S       \label{genfun}
\end{align}
reproducing both the expected one point function from BCFT by differentiating once with respect to the source, and the modified two point function by differentiating twice. Note that for the solutions we have considered, $C_2$ would be set to zero in the Dirichlet case.

The same procedure can be implemented to obtain the generating functional for the ETW brane in the black hole background. For the case of zero-tension BTZ there is a shortcut simply by performing a coordinate transformation which replaces the correlation functions in \ref{genfun} with \ref{btzcorrelators1} and \ref{btzcorrelators2}.

\section{Escaping Boundary State Black Holes}

We show in this section how a microstate dependent deformation of the Hamiltonian can inject negative stress energy into the black hole causing the horizon to recede inwards. This will be demonstrated in two ways: the first by computing the contribution to the average stress energy on the horizon from the Hamiltonian deformation and showing it to be negative, and the second by finding an enhancement of a commutator between early and late operators coming from signals escaping the black hole horizon. For simplicity, we will focus on states where the one point function coefficients are set to zero.

\subsection{A State-dependent Hamiltonian Deformation}

Just as in \citep{Gao:2016bin} we consider a relevant deformation to allow for control over the calculation. The deformation we consider  is
\begin{align}
\delta H_g (t) = g \int d^{d-1} \vec{x} :\mO^2(t, \vec{x}): \label{deform}
\end{align}
where by normal ordering we mean
\begin{align}
:\mO^2(X): = \lim_{X' \rightarrow X}\left[ \mO(X) \mO(X') - \lan \beta |   \mO(X) \mO(X')        | \beta \ran \right]
\end{align}
As we will see later on, the sign of $g$ that produces the desired effect has to be tuned to the microstate of the black hole. We will also allow for the sum of such terms for different species in the case where the bulk contains many fields. Again, the coupling for each term has to be tuned to the specific brane microstate (the boundary condition of the specific bulk field on the brane).

\subsection{Probes of Escapability}\label{Probes}

\subsubsection{Negative Stress Tensor on the Horizon}\label{NegativeStressTensor}

\begin{figure}
\begin{center}
\includegraphics[height=3.5cm]{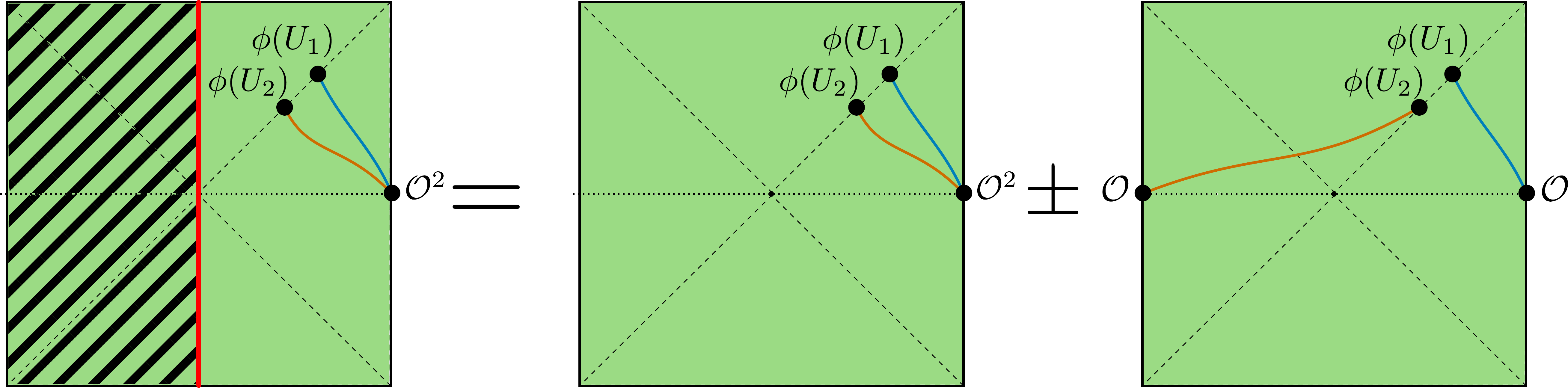}
\caption{The point-split correlation function in the B-State black hole can be expressed as a sum/difference of a correlation functions in the BTZ background. The blue corresponds to a retarded Green's function, while the orange corresponds to a Feynman's Green's function. The first contribution on the right side does not contribute to the integral of the stress tensor on the horizon. The second term is precisely that of the traversable wormhole.}\label{BStateasBTZ}
\end{center}
\end{figure}

For our first probe, we want to compute the contribution to the stress tensor on the future horizon\footnote{Note that we have other contributions to the background stress tensor on the horizon coming from the modified form of the two point function in the boundary state that might displace the horizon slightly outwards. This doesn't really interfere with our effect; the black hole would be escapable from this new deformed horizon.} of the black hole coming from the deformation \ref{deform}. Just as in the eternal black hole geometry, the integral of the stress tensor along the combined past and future horizon is gauge invariant. Since we are interested in $T_{UU}$, where $U$ is the usual outgoing Kruskal coordinate, we can express it via point splitting as
\begin{align}
T_{UU}(U_1) = \lim_{U_2 \rightarrow U_1} \partial_{U_1} \partial_{U_2} \lan {B_\beta} | \phi(U_1) \phi(U_2) | {B_\beta}\ran
\end{align}
where both fields are restricted to be on the future horizon of the black hole, but initially separated in the $U$ direction (we set $V_1 = V_2 = 0$ and $x_1 = x_2 = 0$ taking advantage of the translational symmetry along the $x$ direction).  The two point function in the interaction picture is
\begin{align}
\lan B_\beta |  \bar{T} e^{i \int_{t_0}^{t_1} \delta H_g(t) dt }  \phi(U_1) T e^{-i \int_{t_0}^{t_1} \delta H_g(t) dt } \bar{T} e^{i \int_{t_0}^{t_2} \delta H_g(t) dt }\phi(U_2) T e^{-i \int_{t_0}^{t_2} \delta H_g(t) dt }    | B_\beta \ran 
\end{align}
Since we are working perturbatively about the black hole background, causality implies that $[ \delta H_g (t), \phi(t_i) ] = 0$ for all $t>t_i$. This means we can extend the integration limits of the integrals in the exponentials to infinity, causing the middle two factors to cancel leaving
\begin{align}
\lan B_\beta |  \bar{T} e^{i \int_{t_0}^{\infty} \delta H_g(t) dt }  \phi(U_1) \phi( U_2) T e^{-i \int_{t_0}^{\infty} \delta H_g(t) dt }    | B_\beta \ran 
\end{align}
Keeping only the first order term in the deformation we are left with
\begin{align}
i \int_{t_0}^{\infty} dt  \lan B_\beta |   [ \delta H_g (t), \phi(U_1) \phi(U_2) ]  & | B_\beta \ran = i g \int_{t_0}^{\infty} dt dx  \lan B_\beta |   [ :\mO^2(t,x):, \phi(U_1) \phi(U_2) ]   | B_\beta \ran
\end{align}
This four point function simplifies by large N factorization to give
\begin{align}
i 2 g \int_{t_0}^{\infty} dt dx  \lan B_\beta |   [ \mO(t,x), \phi(U_1) ]   | B_\beta \ran \lan B_\beta |   \{ \mO(t,x), \phi(U_2) \}   | B_\beta \ran
\end{align}
To compute the stress tensor we have to act with $\partial_{U_1} \partial_{U_2}$ on this expression. This somewhat simplifies the expression to
\begin{align}
2 i g \int_{t_0}^{\infty} dt dx  \left[ \left( \partial_U \lan B_\beta |    \mO(t,x) \phi(U)    | B_\beta \ran \right)^2  - \left( \partial_U \lan B_\beta |  \phi(U)    \mO(t,x)    | B_\beta \ran \right)^2 \right]   
\end{align}
The bulk to boundary propagator for spacelike separation is given by
\begin{align}
\lan \phi(r,t_1,x_1) \mO(t_2, x_2) \ran_{B_\beta} &= {r_+^\Delta \over 2^{\Delta + 1} \pi} \left[  - {\left( r^2 - r_+^2\right)^{1/2} \over r_+} \cosh\left[  {2 \pi \over \beta} (t_1 - t_2)  \right]  + {r \over r_+} \cosh\left[  {2 \pi \over \beta} (x_1 - x_2)  \right]  \right]^{-\Delta} \nonumber \\
& \pm {r_+^\Delta \over 2^{\Delta + 1} \pi}  \left[ {\left( r^2  - r_+^2\right)^{1/2} \over r_+}  \cosh\left[  {2 \pi \over \beta} (t_1 + t_2)  \right]  + {r \over r_+} \cosh\left[  {2 \pi \over \beta} (x_1 - x_2)  \right]  \right]^{-\Delta} \nonumber \\
&\equiv K(r,t_1,x_1; t_2, x_2) \pm K(r,t_1,x_1; -t_2 + i \beta/2, x_2)
\end{align}
Where the sign depends on the boundary condition of the field on the brane. Note that these are just bulk to boundary propagators in the eternal black hole hole background. The first term is a right-right propagator while the second is a right-left propagator. For timelike separations, the first term picks up a phase:
\begin{align}
\lan \phi(r,t_1,x_1) \mO(t_2, x_2) \ran_{B_\beta} = e^{-i \pi \Delta}K(r,t_1,x_1; t_2, x_2) \pm K(r,t_1,x_1; -t_2 + i \beta/2, x_2)
\end{align}
The other ordering for the operators picks a different phase giving
\begin{align}
\lan \mO(t_2, x_2) \phi(r,t_1,x_1)  \ran_{B_\beta} = e^{i \pi \Delta}K(r,t_1,x_1; t_2, x_2) \pm K(r,t_1,x_1; -t_2 + i \beta/2, x_2)
\end{align}
Taking the $U_1$ derivative, squaring, and taking the difference of these two orderings we get,
\begin{align}\label{TensorCommutator}
i \int_{t_0}^{\infty} dt  \lan B_\beta |   [ \delta H_g (t), \phi(U_1) \phi(U_2) ] | B_\beta \ran &= -4  g \sin[2 \pi \Delta] \left( \partial_{U_1} K(U_1,V_1, x_1; t_2, x_2) \right)^2  \nonumber \\ 
 \mp 8  g \sin [\pi \Delta]   \partial_{U_1}  K(U_1,   V_1 & ,  x_1;  t_2, x_2) \partial_{U_1} K(U_1,V_1, x_1; -t_2 + i {\beta \over 2}, x_2) 
\end{align}
We find that the contribution to the stress tensor from the first term above vanishes\footnote{In fact had it not vanished, this deformation would be able to make a wormhole traversable via a single sided deformation, leading to paradoxes.}, therefore we restrict our focus on the second term. The change in the stress tensor due to this term is
\begin{align}
\delta T_{UU}(U) = \mp 8g  \sin[\pi \Delta]  \int_{t_0}^\infty dt_2 dx \partial_{U} K(U,0, 0; t_2, x) \partial_{U} K(U,0, 0; -t_2 + i {\beta \over 2}, x) 
\end{align}
The propagator in Kruskal coordinates is
\begin{align}
K(U_1, V_1, x_1; t_2, x_2) & = {r_+^\Delta \over 2^{\Delta + 1} \pi}  \left( {1 + U_1 V_1 \over U_1/U_2 - U_2 V_1  - (1 - U_1 V_1)\cosh[{2 \pi \over \beta} (x_1 - x_2)]} \! \right)^{\Delta} \\
K(U_1, V_1, x_1; -t_2 + i {\beta \over 2}, x_2) &=  {r_+^\Delta \over 2^{\Delta + 1} \pi}  \left(  {1 + U_1 V_1 \over U_1U_2 -  V_1/U_2  + (1 - U_1 V_1)\cosh[{2 \pi \over \beta} (x_1 - x_2)]} \right)^{\Delta}
\end{align}
where $U_2 = e^{2 \pi t_2/\beta}$. Finally, the stress tensor contribution we get is
\begin{align}
\delta T_{UU}(U) \propto \pm g \int_{U_0}^\infty {dU_2 \over U_2} \int_{1}^{U/U_2} {dy \over \sqrt{y^2 - 1}} \partial_U \left[ {U_2 \over U - U_2 y}\right]^\Delta \partial_U \left[ {1 \over U U_2 + y}\right]^\Delta
\end{align}
which is exactly what was found in \cite{Gao:2016bin}. Therefore, we find that escapability of these B-state black holes follows from the traversability of wormholes. All we need to do is choose the correct sign of $g$ to ensure a negative integral of $\delta T_{UU}$. This calculation is presented in figure \ref{BStateasBTZ}.

\subsubsection{Early/Late Time Commutator}

A more direct probe of escapability is to see whether a particle thrown into the black hole can make it back out to the exterior. This can be diagnosed by computing the commutator between an operator inserted at early times, before turning on the deformation, creating a particle which falls in the black hole and another operator at late times around where we expect the particle to come out.

\begin{figure}
\begin{center}
\includegraphics[height=4cm]{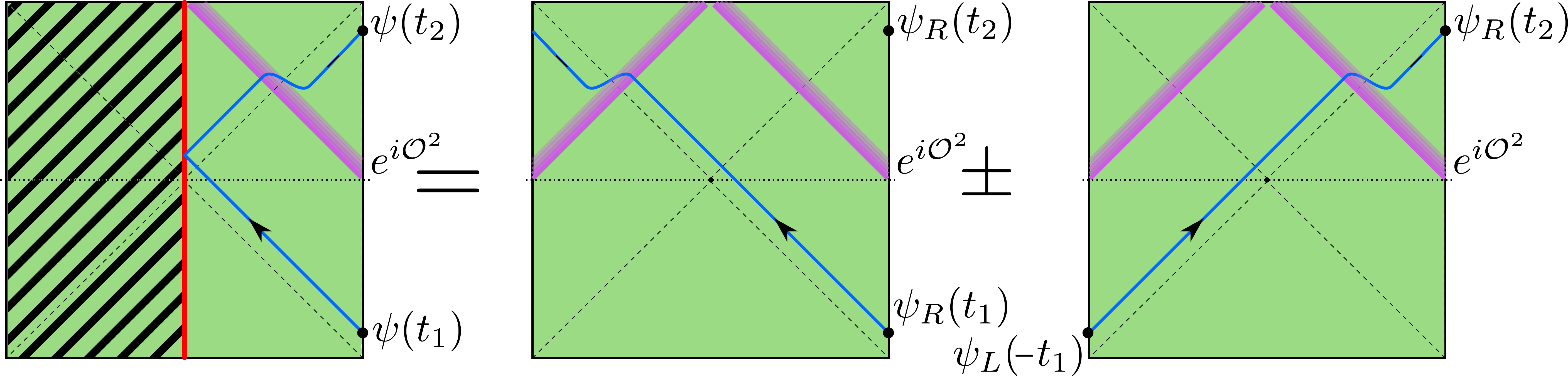}
\caption{The commutator between $\psi(t_1)$ and $\psi(t_2)$ is expected to be significantly enhanced due to turning on the deformation.}\label{commutator}
\end{center}
\end{figure}

Denoting this operator as $\psi$, we wish to compute the effect of the deformation on the commutator
\begin{align}
\lan B_\beta | \left[ \psi(t_1, x), \psi(t_2,x)    \right] | B_\beta \ran.
\end{align}
where the times should be thought of as Heisenberg picture times. Let's first understand this quantity before turning on the deformation. We can express the two terms appearing in the commutator as correlation functions in an eternal black hole background:
\begin{align}
\lan B_\beta | \psi(t_1,x) \psi(t_2,x )| B_\beta \ran &= \lan \beta | \psi_R(t_1,x) \psi_R(t_2,x )| \beta \ran \pm \lan \beta | \psi_R(t_1,x) \psi_R(-t_2 + i \beta/2,x )| \beta \ran \nonumber \\
 &= \lan \beta | \psi_R(t_1,x) \psi_R(t_2,x )| \beta \ran \pm \lan \beta | \psi_R(t_1,x) \psi_L(-t_2,x )| \beta \ran \\
\lan B_\beta | \psi(t_2,x) \psi(t_1,x )| B_\beta \ran &= \lan \beta | \psi_R(t_2,x) \psi_R(t_1,x )| \beta \ran \pm \lan \beta | \psi_R(t_2,x) \psi_R(-t_1 + i \beta/2,x )| \beta \ran \nonumber  \\
 &= \lan \beta | \psi_R(t_2,x) \psi_R(t_1,x )| \beta \ran \pm \lan \beta | \psi_R(t_2,x) \psi_L(-t_1,x )| \beta \ran
\end{align}
Noting that 
\begin{align}
\lan \beta | \psi_R(t_1,x) \psi_L(-t_2,x )| \beta \ran = \lan \beta | \psi_L(-t_1,x) \psi_R(t_2,x )| \beta \ran
\end{align}
we find the commutator to be
\begin{align}
\lan B_\beta | \left[ \psi(t_1, x), \psi(t_2,x)    \right] | B_\beta \ran = \lan \beta | \left[ \psi_R(t_1, x), \psi_R(t_2,x)    \right] | \beta \ran \pm \lan \beta | \left[ \psi_L(-t_1, x), \psi_R(t_2,x)    \right] | \beta \ran \label{thecommutator}
\end{align}
Just as in the previous section, the question of escapability has been reformulated as that of traversablility of the wormhole. From figure \ref{commutator}, it is clear that the first term will not be significantly enhanced by the deformation, but will nevertheless be non-vanishing since $\psi_R(t_2)$ is to the future of the infalling signal. At late times, when $t_2 - t_1 \sim O(t_{scrambling})$, it will be suppressed by $G_N$. The second term is the probe of traversability discussed in \cite{Maldacena:2017axo}, which should be non-vanishing in the presence of the deformation.

Both terms can obtained from one another through analytic continuation. Let's begin by focusing on the left-right commutator, working in the interaction picture making the deformation manifest:
\begin{align}
\lan \beta | \left[ \psi_L(-t_1, x_1), \psi_R(t_2,x_2)    \right] | \beta \ran = -\mathrm{Im} \lan \beta | e^{-i  \int_0^t dt' \delta H(t')}  \psi_R(t_2,x_2) e^{i   \int_0^t dt' \delta H(t')} \psi_L(-t_1,x_1)  | \beta \ran
\end{align}
where the deformation is given by
\begin{equation}
\delta H(t') =   \frac{g}{K}  \sum_{i = 1}^K \int dx'  \mO^2_i(t', x')
\end{equation}
where, just as in \cite{Maldacena:2017axo}, we assume a large number of light fields in order to simplify the calculation. Moreover, since we have mapped the problem to one in the eternal black hole, the deformation really should be thought of as
\begin{align}
\delta H(t') =   \frac{g}{K}  \sum_{i = 1}^K \int dx'  \mO^i_L(-t', x') \mO^i_R(t',x')
\end{align}
At large $K$ we can write
\begin{align}\label{quantity}
e^{-i  \int_0^t dt' \lan \beta |  \delta H(t') | \beta  \ran} \lan \beta | \psi_R(t_2,x_2) e^{i   \int_0^t dt' \delta H(t')} \psi_L(-t_1,x_1)  | \beta \ran 
\end{align}
Using the techniques in \cite{Shenker:2014cwa}, the rest of this correlation function can be evaluated as a scattering between the bulk wave functions sourced by boundary operators appearing in the correlation function. The details of the calculation are presented in appendix \ref{comm}.
The calculation is done in the probe limit, where we ignore the backreaction of the bulk fields on the geometry. This requires that the inequality $G_N p e^t \ll 1$ to hold for the saddle point momentum $p$ of the scattering quanta. As we will see later, there is a window of time where the black hole becomes traversable before backreaction has to be considered.

\begin{figure}
\begin{center}
\includegraphics[height=5cm]{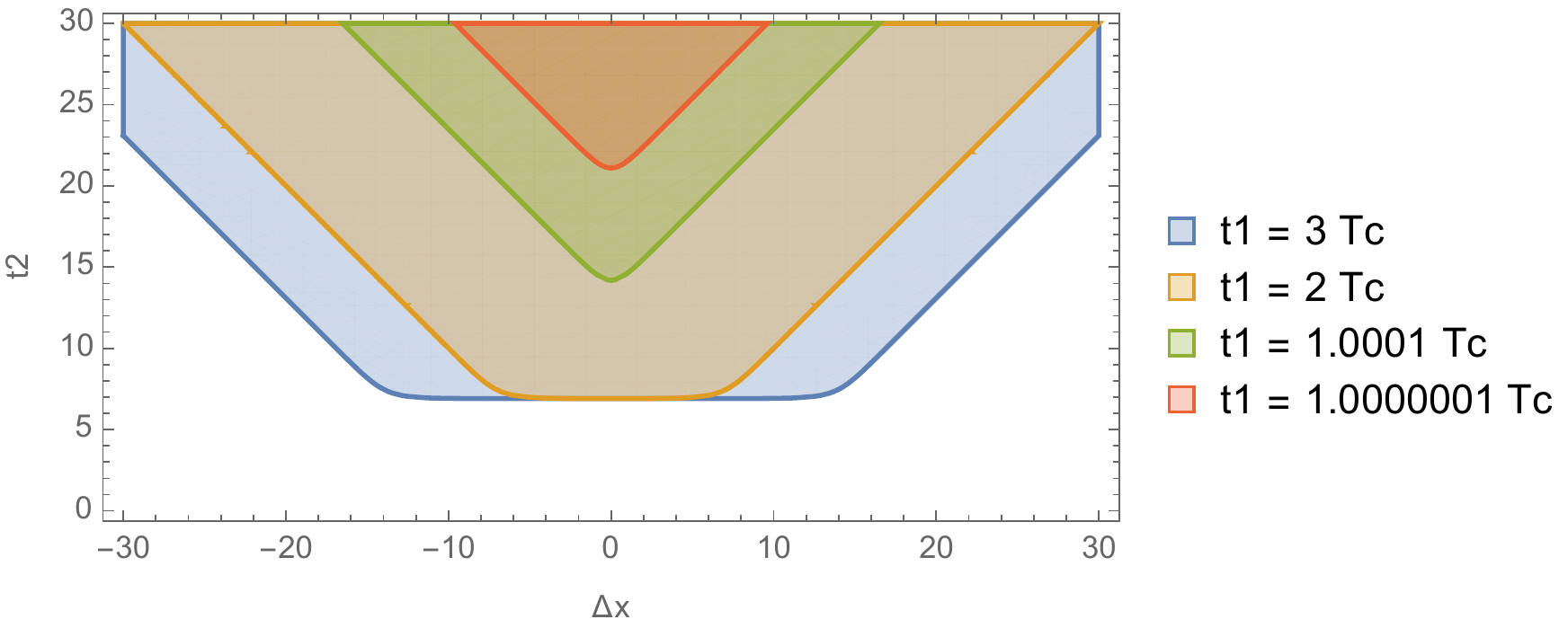}
\caption{This plot shows the domain of nonzero commutator (shaded regions) as a function of $t_2$, the time coordinate of the future probe, and $\Delta x$, the relative spatial displacement of the past and future probes, as the inital probe insertion time, $t_1<0$, is varied. As the magnitude of $t_1$ is decreased (particle is sent from a more recent past), the domain of the nonzero commutator shrinks and gets pushed to future times. }\label{t1dxt3}
\end{center}
\end{figure}

One can study the quantity \ref{quantity} in many limits. The easiest case to analyze is when the deformation lasts for only an instant of time, $\delta H(t') \propto \delta(t' - t)$. In this situation we find that the correlation function is proportional to the integral
\begin{align}
\propto \int d\tilde{x} \left( e^{-t_2} \cosh(\tilde{x} - \Delta x/2)  + e^{t_1} \cosh(\tilde{x} + \Delta x/2) + {g G_N a_1  \over \left[\cosh(t)\right]^{2 \Delta_{\mO} +1}}  \pm i \epsilon \right)^{-2 \Delta_\psi}
\end{align}
where $a_1$ is function of $\Delta_{\mO}$, and the sign of $i \epsilon$ determines the ordering. From this integral we can extract where the commutator will be non-zero; we need to find the region where the denominator vanishes. This will only happen if $g<0$. This singularity signifies the configuration where the two probes, $\psi(t_1)$ and $\psi(t_2)$, are light-like related to one another. We find this region to be
\begin{align}
\cosh[t_2 + t_1] + \cosh[\Delta x] - {K^2 \over 2}e^{t_2 - t_1} \le 0 \label{region}
\end{align}
where
\begin{align}
K = - {g G_N a_1  \over \left[\cosh(t)\right]^{2 \Delta_{\mO} +1}}
\end{align}

\begin{figure}
\begin{center}
\includegraphics[height=5cm]{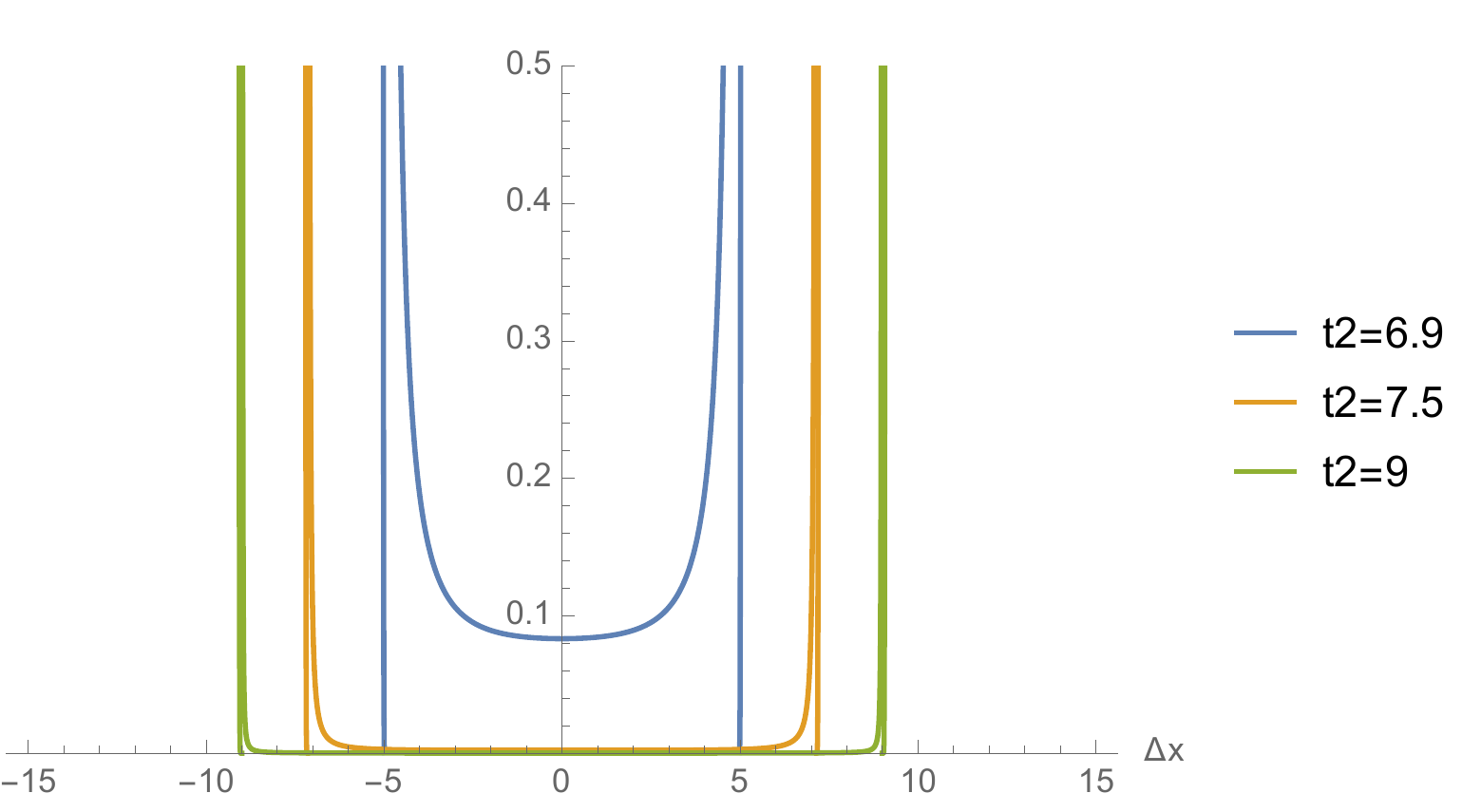}
\caption{This shows the profile of the commutator for $t_1 = 2 T_c$ on constant $t_2$ slices as a function of $\Delta x$. Note how the commutator vanishes more rapidly as $t_2$ is increased.}\label{Constant t1}
\end{center}
\end{figure}

This region is plotted in figure \ref{t1dxt3}. One can show from \ref{region} that there is a critical time, $T_c = \log K <0 $, after which the inserted probe cannot escape the black hole; as $t_1 \rightarrow T_c$ form below, $t_2 \rightarrow \infty$. We expect for earlier (more negative) insertion times that more of the probe wave function makes it outside the black hole leading to a larger spatial region for the commutator, and indeed this is reproduced in the figure. For later times, we expect the portion of the wave function that makes it outside the horizon to asymptote to a point making the region of non-vanishing commutator that of the light cone of a point outside the event horizon of the black hole.
 
\begin{figure}
\begin{center}
\includegraphics[height=5cm]{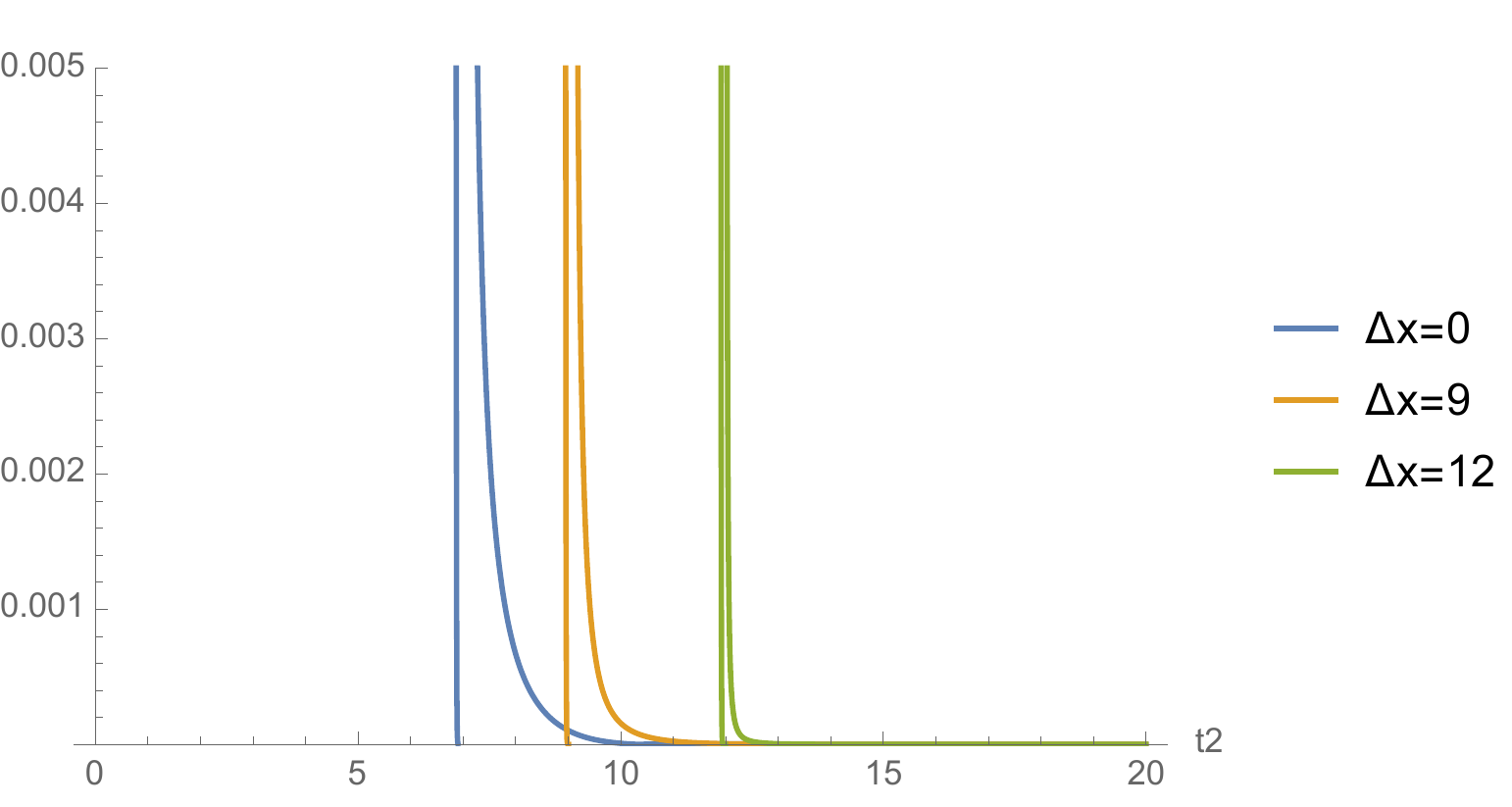}
\caption{This shows the profile of the commutator for $t_1 = 2 T_c$ on constant $\Delta x$ slices as a function of $t_2$. Again, the decay rate is faster for larger $\Delta x$, and effectively, larger $t_2$.  }\label{Constant x}
\end{center}
\end{figure}

Moreover, in this probe approximation both the correlation function and the commutator diverge when the points are light like separated, i.e. on the boundary of the regions presented in figure \ref{t1dxt3}. Both also decay exponentially fast away from the boundaries, and at a faster rate as $t_2$ is increased. This behavior is plotted in figures \ref{Constant t1} and \ref{Constant x}. We leave a more detailed analysis of the commutator, including integrating the deformation, for appendix \ref{comm}. 

Note that the critical time $T_c$ is comparable to the scrambling time, which is at tension with the probe approximation. In particular, if we turn on the deformation at $t = 0$ we can write $T_c \sim -t_{scr} + \log(\-ga_1)$. Thus, if we wish for the probe approximation to hold, we require the inequality $G_N p e^t \ll 1$ to hold for a finite window of time between $-t_{scr}$ and $T_c$. Before the scrambling time, the momenta $p$ that contribute are independent of $g$ and of the thermal scale. This leaves a window of size $\delta t \sim \log(-ga_1)$ where the commutator is non-vanishing while we still remain in the probe approximation (Note that this means that the times used for $t_1 = 2,3 T_c$ are outside the probe approximation, but perhaps one can consider a probe with a momentum profile centered about a small $p\sim G_N$ to keep the approximation valid). This shows the importance of having a large, $O(1)$ value for $g$ in order to recover particles: if $g$ was perturbatively small then $\text{log}(-ga_1) < 0$ and the probe approximation would be invalid before we obtain a non-vanishing commutator. This window corresponds, in Kruskal coordinates, to $\delta U \sim g G_N$. The $G_N$ factor comes from $T_c \sim -t_{scr}$. The quanta emitted in this window will reflect off the ETW brane and stay within $\delta V \sim g G_N$ on the horizon, before encountering the deformation and escaping. This agrees with the shift $\delta V \sim G_N \int dU T_{UU} \sim g G_N$ which comes from the negative stress energy tensor. These findings parallel those of \cite{Maldacena:2017axo}.

We do not investigate this effect outside the probe approximation, but again we expect features similar to those found in \cite{Maldacena:2017axo}. In particular, starting with $t_2 \sim  - t_1 \sim  t_{scr}$, we find that the first term of eq \ref{thecommutator} is subleading in $G_N$, and remains so as $t_2$ or $-t_1$ are increased. Note that this term  would become large as $t_2 \rightarrow t_1$. The second term  of \ref{thecommutator} gets modified in two ways: the backreaction would regulate the commutator stopping it from diverging as found in the probe approximation, and it will be non-zero for all $t_2>t>t_1$, where $t$ is the time where the deformation is first turned on. The latter property would follow because the left probe could interact gravitationally with the left part of the coupling which would instaneously transmit the effect to the right CFT.

\subsubsection{Generalization to Different B-states}

The calculation we considered in the previous sections can be applied to more general states. We consider single-sided black holes which are obtained by cutting out part of the BTZ spacetime behind the black hole horizon and imposing Dirichlet/Neumann boundary conditions on the ETW brane. For concreteness, we can consider the states with non-zero tension ETW branes, whose trajectories are described by \ref{LorentzianBrane}. Using the geodesic approximation, we can write correlation functions for these states as
\begin{equation}
\langle \mO(x,t) \mO(x',t') \rangle = \langle \mO(x,t) \mO(x',t') \rangle_{BTZ} \pm \langle \mO(x,t) \mO(x',t')\rangle_{image}
\end{equation}
Here, the image contribution comes from the geodesic that starts from $(x,t)$, hits the ETW brane and then reaches $(x', t')$. The ETW brane acts as a ``moving mirror", reflecting geodesics in a manner dependent on its velocity. The exact dynamics can be figured out by switching to a frame where the brane is stationary at the point of collision, then the geodesic suffers the familiar ``equal angle" reflection and we can boost back to the original frame.  

While calculating the length of this geodesic is difficult in general, we can argue that it will be an $O(1)$ number (in the $1/N$ expansion) as long as $t, t'$ are $O(1)$. The reason is that if we look at figure \ref{BHBrane}, we expect the geodesic to lie entirely away from the ``corners" of the diagram. The metric has $O(1)$ components away from the corners and the singularity, and besides for a UV contribution near the boundary (which we cut off anyways) the geodesic should have $O(1)$ coordinates. These two conditions give an $O(1)$ length, so the geodesic approximation suggests that
\begin{equation}
\langle \mO(x,t) \mO(x',t')\rangle_{image} \sim O(1)
\end{equation}
More generally, we expect that the image contribution to the bulk to boundary propagator will be of the same order (once again, assuming we keep the bulk point away from the corners). The same computation as in section \ref{NegativeStressTensor} then applies and it should give an $O(1)$ shift to the stress-energy tensor on the horizon, with a sign that can be chosen to be negative. This may be cumbersome to do explicitly due to the complicated forms of the geodesics, but barring a bizarre cancellation the argument should follow through for general values of the brane tension. After all, the crucial component of the computation wasn't the exact form of the bulk to boundary propagator, but instead the fact that the image term was $O(1)$ and real. The latter condition is necessary to ensure that there won't be additional contributions to the commutator in \ref{TensorCommutator} and it follows from the fact that the image geodesic is spacelike.

\section{Discussion}

\subsection{Microstate Dependence}

We found in the preceding sections that the deformation that makes the black hole escapable has to be tailored to the black hole microstate in question. We argue here that this is a necessary part of achieving escapability in more general black holes. In particular, only a state dependent deformation can violate the ANEC along a black hole horizon.

The argument is simple: Let's assume there exists a state independent deformation that causes a violation of the ANEC along the horizon for all microstates within some energy band. Now consider a thermofield double state with average energy given by the set of states just considered, and turn on the state independent deformation on, say, the right CFT. The integral of the stress tensor along the horizon of the TFD will simply be the thermal average of that in the individual microstates, and therefore will also be negative. Thus the ANEC will be violated along the horizon of the eternal black hole, transforming it into a traversable wormhole. This would allow a signal to be sent from the left CFT to the right, implying a non-trivial commutator between left and right operators. However, quantum mechanically this cannot follow since the two CFTs remain decoupled; the deformation is purely right sided.

\subsection{Escaping at Late Times}

In the examples considered in this paper, escapability was only possible at early times due to the exponential drop-off of the one point function $\lan B | \mO^2(t) | B \ran$. The state essentially thermalizes and one point functions of simple operators vanish. This thermalization precludes the existence of a simple deformation that could induce escapability. 

However, it does leave the possibility of complicated deformations, even more fine tuned to the state than what we considered, that might get the job done. Understanding more details of boundary states would be required to achieve this, and perhaps doable in the context of SYK \cite{Kourkoulou:2017zaj}. Nevertheless, there still remains the worry that the fine tuning of the deformation to the state might be so sensitive that it immediately fails once an object/message is thrown into the black hole.


\subsection{The Deformed ADM Energy}

Here we repeat the calculation of the effect of the deformation on the ADM energy already discussed somewhat in \cite{Gao:2016bin, Maldacena:2017axo}, giving the leading term in the large $K$ limit and the first subleading term in the $1/N$ expansion. For simplicity we consider the case where the deformation is turned on for an instant of time. Consider computing the expectation value of the undeformed Hamiltonian in the interaction picture,
\begin{align}
\lan B | e^{i \delta H(t)} H_0 e^{- i \delta H(t)} | B \ran
\end{align}
If we make the replacement $e^{- i \delta H(t)} | B \ran = | B \ran  e^{- i \lan \delta H(t) \ran_B}$ we would find that
\begin{align}
\lan B | e^{i \delta H(t)} H_0 e^{- i \delta H(t)} | B \ran = \lan B |  H_0 | B \ran 
\end{align}
and therefore the energy does not change. However, this replacement is true only up to subleading corrections in $G_N$ which could get enhanced by the Hamiltonian to an order one correction to the energy, since it has a $1/G_N$ expectation value in this state.

This correction is actually calculable by first commuting the Hamiltonian through the deformation and then making the replacement,
\begin{align}
\lan B | e^{i \delta H(t)} H_0 e^{- i \delta H(t)} | B \ran &= \lan B |  H_0 | B \ran  -  \lan B | e^{i \delta H(t)}  \delta H'(t)  e^{- i \delta H(t)}  | B \ran \\
&= \lan B |  H_0 | B \ran  -  \lan B |  \delta H'(t)   | B \ran + \mO(G_N)
\end{align}
Here we relied on the expectation value of $\delta H'(t)$ being order one and therefore not enhancing the error from the replacement of $\delta H$ in the exponent with its expectation value. The change in energy is then
\begin{align}
\Delta E &= - {g \over K}  \sum_{i = 1}^K   \int dx \partial_t \lan B |  :(\mO^i(t,x))^2: | B \ran \\
&= - {g \over K}  \sum_{i = 1}^K  \int dx \partial_t \lan \beta |  \mO^i_L(-t,x) \mO^i_R(t,x) | \beta \ran \\
&=  {g V_x\over K}  \sum_{i = 1}^K   2 \Delta_i {\sinh[2 t] \over \cosh^{\Delta_i + 1}(2 t)}
\end{align}
where time here is measured in units of $\beta$, and $V_x$ is the volume of the transverse space assumed to be compactified. Just as in the previous work \cite{Gao:2016bin, Maldacena:2017axo}, we see that escapability is uncorrelated with the change in the total ADM energy; the change in energy depends on the sign of $t$, both of which lead to the black hole being escapable.




\subsection{Interior Operators and State Dependence}

\begin{figure}
\begin{center}
\includegraphics[height=5cm]{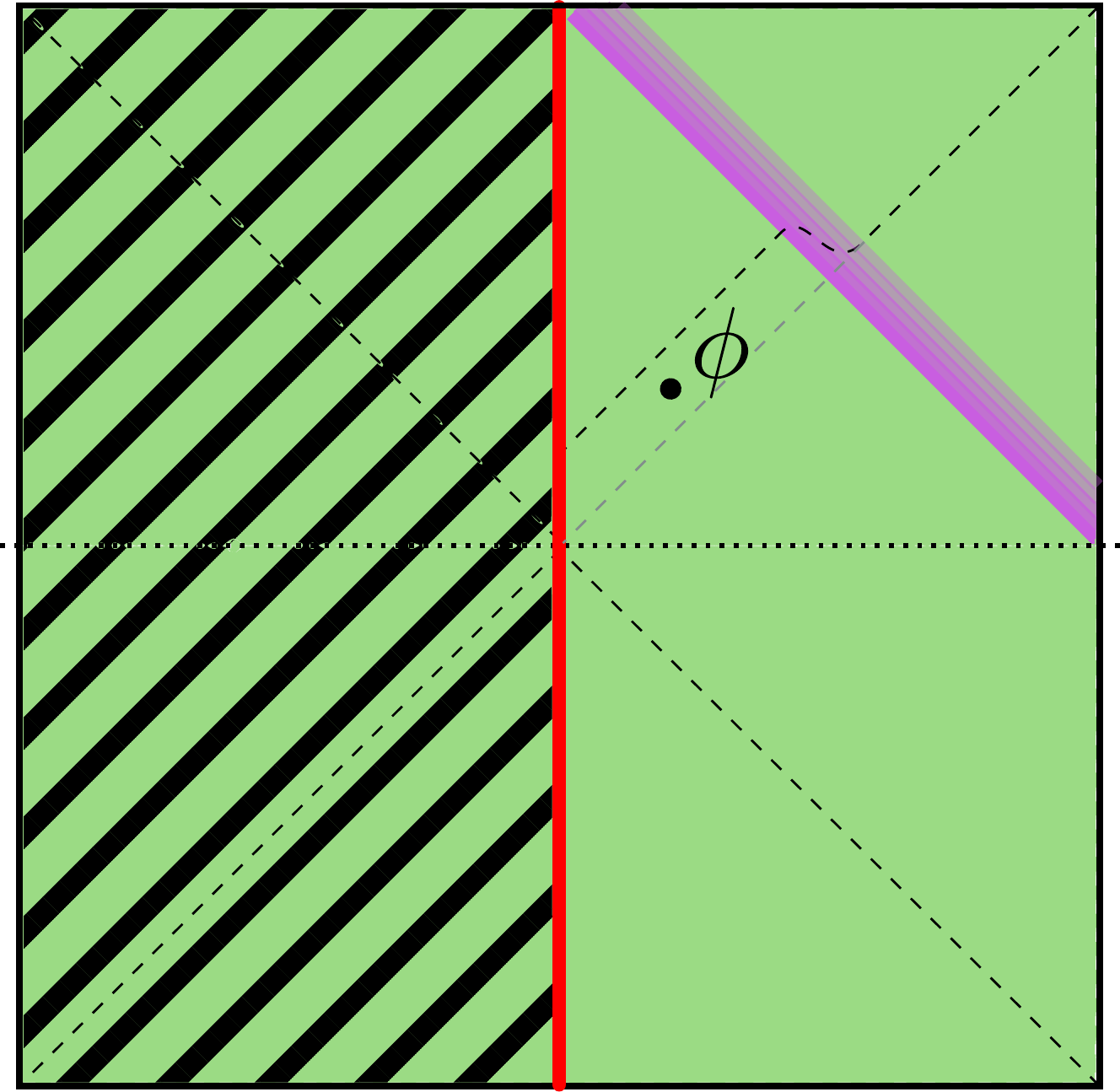}
\caption{The modified black hole spacetime due the deformation. The operator $\phi$ is causally connected to the boundary of AdS. The greyed out dotted line is the location of the undeformed horizon.}\label{Interior Operator}
\end{center}
\end{figure}

Making a black hole escapable gives us a window into the reconstruction of operators behind its event horizon, particularly via the HKLL prescription. Consider the setup presented in figure \ref{Interior Operator}. In the original Boundary state black hole spacetime, the operator $\phi$ lies within the interior of the black hole and, therefore, the usual HKLL prescription of solving the bulk wave equations to represent the operator using a spacelike green's function won't work due to part of the singularity being spacelike separated to $\phi$. With the inclusion of the deformation, however, the spacetime gets modified and part of the original black hole interior becomes in causal contact with the boundary, allowing for the usual HKLL prescription.

This can be done by following the steps outlined in \cite{Almheiri:2017fbd}. Consider first the undeformed state $| B \ran$ dual to the regular un-escapable boundary state black hole. We imagine working in a gauge defined by starting with points on the brane and going out towards the boundary in a spacelike direction, specifying each bulk slice with a boundary time.  Let's now posit the existence of an operator $\phi_{in}$ that creates an excitation inside the black hole event horizon. Let's say this operator acts on the CFT at some time $t>0$:
\begin{align}
\phi_{in}^t e^{- i H_0 t} | B \ran
\end{align}
where the superscript $t$ indicates the boundary slice we are working on. The idea is that now we can evolve the state back in time and turn on the deformation which makes the location of this operator outside the black hole, and evolve back to the original time:
\begin{align}
{\cal T}e^{- i \int_{0}^{t} H(t') dt'}  e^{i H_0 t}\phi^t_{in} e^{- i H_0 t} | B \ran
\end{align}
where $H = H_0 + \delta H(t)$, and the deformation has some explicit time dependence. Note, we assume that our original choice of gauge continues to be good (without caustics) in the deformed spacetime. Now, since the deformation renders the location of the operator in causal contact with the boundary, this implies that there is an HKLL prescription, but in the deformed geometry, for representing the excitation as smeared boundary operators, $\phi_{HKLL}^t$. Therefore, the above state could be rewritten as
\begin{align}
\phi_{HKLL}^t {\cal T}e^{- i \int_{0}^{t} H(t') dt'}  | B \ran 
\end{align}
We stress that it is understood how to find $\phi_{HKLL}^t$ in the new geometry. Finally, we just have to equate the two expressions to obtain an explicit form of the operator behind the horizon as
\begin{align}
  \phi_{in}^t &=  e^{-i H_0 t} \  \bar{{\cal T}}e^{ i \int_{0}^{t} H(t') dt'} \phi_{HKLL}^t {\cal T}e^{- i \int_{0}^{t} H(t') dt'} e^{i H_0 t} \\
  &= \bar{{\cal T}}e^{ i \int_{0}^{t} \delta H(t') dt'} \phi_{HKLL}^t {\cal T}e^{- i \int_{0}^{t} \delta H(t') dt'}
\end{align}
This result warrants a few comments. This interior operator depends on $\delta H$, which has to be fine tuned to the state in question to render the black hole escapable. Therefore, $\phi^t_{in}$ is also state dependent. Note that this is more than the usual notion of background dependence, discussed in \cite{Heemskerk:2012mn}, since the different boundary state black holes have the same backgrounds.

An interesting question here is whether this construction gets around the trans-Planckian problem of reconstructing late modes behind the black hole horizon. The problem is the following: Consider the case of a boundary state black hole and focus on a late `out-going' mode behind the horizon. Working in the probe approximation it seems that this mode can be simply evolved backwards, reflecting of  the brane, evolved all the way outside the black hole where we can use the usual HKLL prescription. This is too quick, however, since the large relative boost between the mode and the brane would likely cause the formation of a new black hole horizon shielding the mode from being reconstructable using HKLL.

However, it is clear from the new spacetime with the deformation that one can evolve the data of the late mode to a region that is causally connected to the boundary without reflecting of the brane! The problem with this picture is that the causal structure depends on the relative boost of the modes and the negative energy flux; the stronger the collision between the two the smaller the amount of time advance of the modes \cite{Maldacena:2017axo}. This means that the evolved back late modes are not actually causally connected to the boundary precluding the HKLL prescription. The trans-Planckian problem persists.

\acknowledgments
It's a pleasure to thank Daniel Harlow, Thomas Hartman, Daniel Jafferis, Juan Maldacena, Eva Silverstein, Douglas Stanford, and Aron Wall for useful input and conversations. MS is supported in part by the NSF grant PHY-1316699. AM is supported by the Stanford Graduate Fellowship.
\appendix

\section{Details of Early/Late time commutator\label{comm}}

In this appendix, we would like to calculate the \textit{early-late} commutator, and use it as a probe for escapability. The primary details of the calculation are based on the shockwave results of \cite{Shenker:2014cwa}, and the arguments about traversability in \cite{Maldacena:2017axo}. As shown in the main text the commutator in the boundary state is given by,
\begin{align}
\lan B_\beta | \left[ \psi(t_1, x), \psi(t_2,x)    \right] | B_\beta \ran = \lan \beta | \left[ \psi_R(t_1, x), \psi_R(t_2,x)    \right] | \beta \ran \pm \lan \beta | \left[ \psi_L(-t_1, x), \psi_R(t_2,x)    \right] | \beta \ran. \nonumber
\end{align}
In the presence of the deformation,
\begin{equation}
\delta H(t') =   \frac{g}{K}  \sum_{i = 1}^K \int dx'  \mO^i_L(t', x') \mO^i_R(-t', x')
\end{equation}
for large $K$, it can be argued \cite{Maldacena:2017axo} that the commutator becomes,
 \begin{align}
C  = & \, e^{-i \int_0^t dt' \lan \delta H(t') \ran } \braket{\beta |\psi_R(t_2,x_2) e^{i \int_0^t dt' \delta H(t') } \psi_R(t_1,x_1)|\beta} \nonumber \\
& \pm e^{-i \int_0^t dt' \lan \delta H(t') \ran } \braket{\beta|\psi_R(t_2,x_2) e^{i \int_0^t dt' \delta H(t') } \psi_L(-t_1,x_1)|\beta}. \label{comm1}
\end{align}
In the above, the operator ordering in the second term guarantees that the scattering between the $\psi$ and $O$ excitation is enhanced. The ordering in the first term is not sensitive to this scattering, and at late times becomes $\braket{\psi_R(t_2) \psi_R(t_1)} \braket{\delta H}$ and is thus exponentially suppressed in $\Delta t = t_2 - t_1$.  Thus up to an overall coefficient and exponentially suppressed corrections, we are interested in finding 
\be
\tilde{C} \equiv \braket{\beta|\psi_R(t_2,x_2) e^{i \int_0^t dt' \delta H(t') } \psi_L(-t_1,x_1)|\beta}.  \label{comm2}
\ee
To understand the physics, let's first expand the exponential to first order in $g$,
\begin{align}
\tilde{C}_1 & =  \frac{ i g }{K} \sum_{j=1}^{K} \int_{0}^{t} \! \! dt' \! \! \int \! \!  dx \braket{\beta|\psi_R(t_2,x_2)  \, O^j_L(-t',x) O^j_R(t',x)  \psi_L(-t_1,x_1)|\beta}  \label{first}
\end{align}
Let's focus on a single term of this sum which can be interpreted as a scattering cross section between the following \textit{in-out} states
\begin{align}
\ket{in} & = \mO_R(t',x)  \psi_L(-t_1,x_1) \ket{\beta} \nonumber \\
\ket{out} & = \psi^\dagger_R(t_2,x_2) \mO^\dagger_L(-t',x)  \ket{\beta} .
\end{align}
The \textit{in} and \textit{out} states can be written in terms of the single particle wave functions as
\begin{align}
& | in \ran = \int \Psi_4(p_4^u, \tilde{x}_4) \Psi_1(p_1^v, \tilde{x}_1)     | p_4^u, \tilde{x}_4; p_1^v, \tilde{x}_1 \ran_{in} \nonumber \\
& | out \ran = \int  \Psi_2(p_2^v, \tilde{x}_2) \Psi_3(p_3^u, \tilde{x}_3)    | p_2^v, \tilde{x}_2; p_3^u, \tilde{x}_3 \ran_{out} \label{states}
\end{align}
where the integral is over all the exposed variables. The $\tilde{x}_i$ is the transverse coordinate, while $p_1^v$ and $p_2^v$ are the ingoing and outgoing momenta of the $\psi$ particle, and $p_4^u$ and $p_3^u$ of the $\mO$ particle. The wave functions are
\begin{align}
&\Psi_4(p_4^u, \tilde{x}_4) = \int dv e^{i a_0 p_4^u v/2} \lan \phi_\mO(u,v,\tilde{x}_4) \mO_R(t^\prime,x^\prime) \ran_{u = 0} \nonumber \\
&\Psi_1(p_1^v, \tilde{x}_1) = \int du e^{i a_0 p_1^v u/2} \lan \phi_\psi(u,v,\tilde{x}_1) \psi_L(-t_1,x_1) \ran_{v = 0}  \nonumber \\
&\Psi_2(p_2^v, \tilde{x}_2) = \int du e^{i a_0 p_2^v u/2} \lan \phi_\psi(u,v,\tilde{x}_2) \psi_R^\dagger(t_2,x_2) \ran_{v = 0} \nonumber \\
&\Psi_3(p_3^u, \tilde{x}_3) = \int dv e^{i a_0 p_3^u v/2} \lan \phi_\mO(u,v,\tilde{x}_3) \mO_L^\dagger(-t^\prime,x^\prime) \ran_{u = 0}.\label{wavefn1}
\end{align}
and the kets $\ket{ p_4^u, \tilde{x}_4}, \, \ket{ p_3^u, \tilde{x}_3} $ and $\ket{ p_1^v, \tilde{x}_1}, \, \ket{ p_2^v, \tilde{x}_2} $ are defined on the Hilbert space on the $u=0$ and $v=0$ slice respectively. Intuitively, we have decomposed the operators $\psi$ and $\mO$ in the basis of longitudinal momentum and transverse coordinates at the horizon, with the help of the bulk (horizon) to boundary two point function. The norm of the position and momentum states is derived from the Klein-Gordon norm, and is
\begin{align}
\lan p, x | q, y \ran = {a_0^2 p \over 4 \pi r_0} \delta(p - q) \delta(x - y).
\end{align}
Plugging in the two point function in \ref{wavefn1},
\begin{align}
\lan \phi(u,v,\tilde{x}) \mO(t,x) \ran  = c_\mO    \left(  {    1 + u v \over u e^{t} - v e^{-t} + (1 - u v)\cosh\left[  \tilde{x} - x  \right]     }      \right)^{2 \Delta}
\end{align}
the wave functions evaluate to
\begin{align}
&\Psi_4(p_4^u, \tilde{x}_4) = \Theta(p_4^u) {2 \pi i c_\mO e^{t'} \over \Gamma(\Delta_\mO)}  \left( {-i a_0 p_4^u e^{t'} \over 2}    \right)^{\Delta_\mO - 1} e^{i {a_0 \over 2} p^u_4 e^{t'}  \cosh\left[  \tilde{x}_4 - x' \right]} \nonumber \\
&\Psi_1(p_1^v, \tilde{x}_1) =\Theta(p_1^v) {2 \pi i c_\psi e^{t_1} \over \Gamma(\Delta_\psi)}  \left( {-i a_0 p_1^v e^{t_1} \over 2}    \right)^{\Delta_\psi - 1} e^{i {a_0 \over 2} p^v_1 e^{t_1}  \cosh\left[  \tilde{x}_1 - x_1 \right]}    \nonumber \\
&\Psi_2(p_2^v, \tilde{x}_2) = \Theta(p_2^v) {2 \pi i c_\psi e^{-t_2^*} \over \Gamma(\Delta_\psi)}  \left( {i a_0 p_2^v e^{-t_2^*} \over 2}    \right)^{\Delta_\psi - 1} e^{-i {a_0 \over 2} p^v_2 e^{-t_2^*}  \cosh\left[  \tilde{x}_2 - x_2 \right]}  \nonumber \\
&\Psi_3(p_3^u, \tilde{x}_3) = \Theta(p_3^u) {2 \pi i c_\mO e^{-t'^*} \over \Gamma(\Delta_\mO)}  \left( {i a_0 p_3^u e^{-t'^*} \over 2}    \right)^{\Delta_\mO - 1} e^{-i {a_0 \over 2} p^u_3 e^{-t'^*}  \cosh\left[  \tilde{x}_3 - x' \right]}.  \label{wavefn2}
\end{align}
The above cross section has a simple interpretation in the bulk: the wave packets produced by $\mO$ and $\psi$ have a large relative boost, as a result the cross section is dominated by the gravitational interaction, which can then be approximated by the gravitational shock-wave amplitude \cite{HOOFT1987,Shenker:2014cwa}. For such high energy scattering, the momentum transfer is very small $t/s \ll 1$, and the transverse co-ordinates are thus approximately conserved. The scattering element between ingoing and outgoing states is:
\begin{align}
{}_{out}\lan p_3^u, \tilde{x}_3; p_2^v, \tilde{x}_2  | p_4^u, \tilde{x}_4 ; p_1^v, \tilde{x}_1 \ran_{in} = \left( {a_0^2  \over 4 \pi r_0} \right)^2 p_1^v p_4^u  e^{i \delta}  \delta(p_1^v - p_2^v) \delta(p_4^u - p_3^u) \delta(\tilde{x}_1 - \tilde{x}_2) \delta(\tilde{x}_4 - \tilde{x}_3) \nonumber
\end{align}
where
\begin{align}
\delta = {2 \pi a_0 G_N \over r_0^2} p_1^v p_4^u e^{-  | \tilde{x}_1 - \tilde{x_4}|}. 
\end{align}
Thus we have,
\begin{align}
\tilde{C}_1 = \int dp^v_1 d\tilde{x}_1 \left[p_1^v  \Psi^*_2(p^v_1,\tilde{x}_1) \Psi_1(p^v_1,\tilde{x}_1) \right] \int dp^u_4 d\tilde{x}_4  \, \left( i \alpha g\right)  \! \! \int dt^\prime dx^\prime \left[p_4^u  e^{i \delta}  \Psi^*_3(p^u_4,\tilde{x}_4)  \Psi_4(p^u_4,\tilde{x}_4)\right] \nonumber
\end{align}
where $\alpha =  \left( {a_0^2  \over 4 \pi r_0} \right)^2$. Using \ref{wavefn2} we can now evaluate $\tilde{C}_1$. At higher orders in $g$ at large $K$, the gravitational scattering continues to dominate exponentially over all the other interactions like self interactions of $\phi$, and $\mO$. Hence the term, $\braket{\beta | \psi_R (\mO \mO)^n \psi_L | \beta}$ at order $O(g^n)$ in \ref{comm2} can be viewed as $n$ separate and independent scattering events \cite{Maldacena:2017axo}. Moreover, we assume that the dimensions of all the $\mO^i$ are the same. Then, adding all the phases and resumming the exponential we have,
\begin{align}
\tilde{C} & = \alpha \! \! \int dp^v_1 d\tilde{x}_1 \left[p_1^v  \Psi^*_2(p^v_1,\tilde{x}_1) \Psi_1(p^v_1,\tilde{x}_1) \right] \text{exp} \left[ i \alpha g \! \!  \int dp^u_4 d\tilde{x}_4  \,  \! \! \int dt^\prime dx^\prime \left[p_4^u  e^{i \delta}  \Psi^*_3(p^u_4,\tilde{x}_4)  \Psi_4(p^u_4,\tilde{x}_4)\right]\right] \nonumber 
\end{align}
Using the wave functions in \ref{wavefn2} and some redefinitions,
\begin{align*}
\tilde{C} =  -  2^{4\Delta_\psi}\alpha b_\psi^2 \int dq & d\tilde{x}_1  q^{2\Delta_\psi -1} e^{i 2 q \left( e^{-t_2}\cosh[\tilde{x}_1-x_2] + e^{t_1} \cosh[\tilde{x}_1 - x_1] \right)}  e^{-i\pi \Delta_\psi} e^{-(t_2 - t_1)\Delta_\psi }\times \nonumber \\ 
\times \text{exp} \bigg  [\! - \! i \alpha g   2^{4\Delta_\mO} b_\mO^2 \! \! \int \! \! & dp d\tilde{x}_4 dt' dx'  \, {p^{2\Delta_\mO - 1}e^{i 4 p  \cosh[\tilde{x}_4-x']\cosh[t']} e^{-i \pi \Delta_\mO }} \text{exp}\left[ \frac{8 \pi i  G_N}{r_0^2} {p q e^{-|\tilde{x}_4-\tilde{x}_1|}} \right] \bigg]
\end{align*}
where 
\begin{align}
b_\psi = \frac{\pi c_\psi}{2^{\Delta_\psi}\Gamma(\Delta_\psi)} ,  \quad b_\mO = \frac{\pi c_\mO}{2^{\Delta_\mO}\Gamma(\Delta_\mO)}.
\end{align}
We will work in the limit of negligible back reaction and assume $G_N p e^ t \ll 1$. Evaluating the integrals in the exponential and expanding it to to first order in $G_N p $ we then find, 
\begin{align}
& \tilde{C} =  - 2^{4\Delta_\psi} \alpha b_\psi^2 \int dq  d\tilde{x}_1  q^{2\Delta_\psi -1} e^{i 2 q \left( e^{-t_2}\cosh[\tilde{x}_1-x_2] + e^{t_1} \cosh[\tilde{x}_1 - x_1] \right)}  e^{-i\pi \Delta_\psi} e^{-(t_2 - t_1)\Delta_\psi }\times \nonumber \\  
& \times \! \text{exp}  \! \left[ \! - i g \! \! \int \! \! dt^\prime \! \left(\! \frac{2b_\mO^2 \alpha L \pi^{\frac 1 2} \Gamma(\Delta_\mO)\Gamma(2\Delta_\mO)}{\Gamma\left(\Delta_\mO + \frac 1 2\right)\cosh(t')^{2\Delta_\mO}} -  q \frac{8  b^2_\mO \alpha G_N \pi^{\frac 3 2} \Gamma\left(\Delta_\mO + \frac 1 2\right)\Gamma(2\Delta_\mO)}{r_0^2\Gamma(\Delta_\mO)\cosh(t')^{2\Delta_\mO + 1}}\! \right) \! \right] \label{noback}
\end{align}
where we have imposed an IR cutoff $L$ on the spatial integral $x^\prime$. Before we proceed to evaluate the above, we need to choose a profile for our deformation. We will begin with a $\delta$-function profile $\delta H(t^\prime) \propto \delta(t^\prime-t)$, and present results for a more general profile below. We find,
\begin{align}
\tilde{C} = -\int d\tilde{x}_1\frac{2^{4\Delta_\psi} \alpha b_\psi^2 e^{-(t_2 - t_1)\Delta_\psi} e^{-i a_2 g L \cosh(t)^{-2\Delta_\mO} } \Gamma(2\Delta_\psi)}{\left ( 2 e^{-t_2} \cosh(\tilde{x}_1 - x_1) + 2 e^{t_1}\cosh(\tilde{x}_1 - x_2 )  + \frac{g G_N a_1}{ \cosh(t)^{2\Delta_\mO +1}} \right)^{2\Delta_\psi}} \label{x2neqx1}
\end{align}
where
\begin{align}
a_2 =  \frac{2b_\mO^2 \alpha  \pi^{\frac 1 2} \Gamma(\Delta_\mO)\Gamma(2\Delta_\mO)}{\Gamma\left(\Delta_\mO + \frac 1 2\right)}, \quad  a_1=  \frac{8  b^2_\mO \alpha  \pi^{\frac 3 2} \Gamma\left(\Delta_\mO + \frac 1 2\right)\Gamma(2\Delta_\mO)}{r_0^2\Gamma(\Delta_\mO)}.
\end{align}
\begin{figure}
\centering
\includegraphics[width=0.65\textwidth]{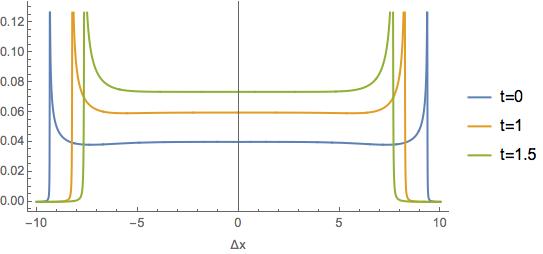}
\caption{$x_2 \neq x_1, \, t_2=|t_1|$ with a delta function source centered around different values of $t$.}
\label{nzt}
\end{figure}
Note that in the limit, $x_2=x_1$ and $t=0$ the above resembles the familiar answer from $AdS_2$,
\begin{align}
\tilde{C} = -\int d\tilde{x}_1\frac{2^{4\Delta_\psi} \alpha b_\psi^2  e^{-i a_2 g L  } \Gamma(2\Delta_\psi)}{ \cosh\left(\frac{|t_1| -t_2}{2}\right)^{2\Delta_\psi} \left ( 4 \cosh(\tilde{x}_1 - x_1) + \frac{a_1 g G_N  e^{\frac{|t_1| + t_2}{2}}}{ \cosh\left(\frac{|t_1| -t_2}{2}\right)} \right)^{2\Delta_\psi}} \label{x2=x1}
\end{align}
For negative $g$, the denominator turns negative at large times and the integral becomes imaginary when,
\begin{align}
\frac{4}{a_1 |g| G_N  }  \le \frac{e^{\frac{|t_1| + t_2}{2}}}{ \cosh\left(\frac{|t_1| -t_2}{2}\right)}.
\end{align}For fixed values of $t_1$ and $\Delta x = |x_1 - x_2|$, we find as in figure \ref{t1dxt3}, that the commutator vanishes for smaller values of $t_2$ as the magnitude of $|t_1|$ is increased. Note that the value of $t_2$ saturates as $|t_1|$ is increased. Similarly, for fixed $t_1$ we find as in figure \ref{Constant t1} that the domain of nonzero commutator decreases as $t_2$ is increased. When the source is turned on for just an instant i.e.  $\delta H(t^\prime) \propto \delta(t' - t)$, we find in figure \ref{nzt} that for later deformations i.e larger $t$, the domain of escapability shrinks. In other words, more of the particle wave function escapes the black hole for smaller $|t|$.
\begin{figure}
\centering
\includegraphics[width=0.65\textwidth]{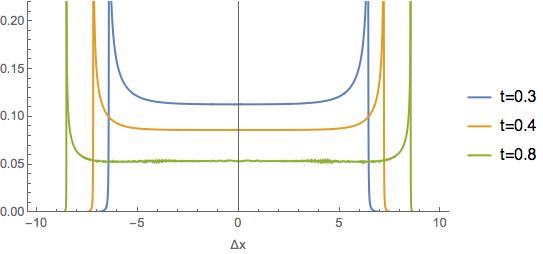}
\caption{$x_2 \neq x_1$ for a source that is turned for time $0$ to $t$ for some $t_1,t_2$.}
\label{fourth}
\end{figure}Finally, we will look at the case when the deformation is not a $\delta$-function, but instead is turned on for some time $\int_0^t dt^\prime$. In this case \ref{noback} becomes
\begin{align}
\tilde{C} = -\int d\tilde{x}_1\frac{2^{4\Delta_\psi} \alpha b_\psi^2 e^{-(t_2 - t_1)\Delta_\psi} e^{-i a_2 g L f\left(t,\Delta_\mO - \frac 1 2\right) } \Gamma(2\Delta_\psi)}{\left ( 2 e^{-t_2} \cosh(\tilde{x}_1 - x_1) + 2 e^{t_1}\cosh(\tilde{x}_1 - x_2 )  + g G_N a_1f(t,\Delta_\mO)  \right)^{2\Delta_\psi}} \label{x2neqx1}
\end{align}
where,
\begin{align}
f(t,\Delta_\mO)=\left.\frac{i}{2}  \left(B_{\cosh ^2(t)}\left(-\Delta_\mO ,\frac{1}{2}\right)-\frac{\sqrt{\pi } \Gamma (-\Delta_\mO )}{\Gamma \left(\frac{1}{2}-\Delta_\mO \right)}\right)\right)
\end{align}
and $B$ represents the incomplete beta function. As can be seen in figure \ref{fourth}, turning on a deformation for longer times makes the domain of escapability larger. Note that $f(t,\Delta)$ saturates to $1$ at large times, thus there is a lower bound on the time it takes for the information to reappear.


\bibliographystyle{JHEP}
\bibliography{bibliography}

\providecommand{\href}[2]{#2}\begingroup\raggedright\begin{thebibliography}{10}

\bibitem{Almheiri:2012rt}
A.~Almheiri, D.~Marolf, J.~Polchinski, and J.~Sully, {\it {Black Holes:
  Complementarity or Firewalls?}},  {\em JHEP} {\bf 02} (2013) 062,
  [\href{http://arxiv.org/abs/1207.3123}{{\tt arXiv:1207.3123}}].

\bibitem{Hawking:1976ra}
S.~W. Hawking, {\it {Breakdown of Predictability in Gravitational Collapse}},
  {\em Phys. Rev.} {\bf D14} (1976) 2460--2473.

\bibitem{Papadodimas:2012aq}
K.~Papadodimas and S.~Raju, {\it {An Infalling Observer in AdS/CFT}},  {\em
  JHEP} {\bf 10} (2013) 212, [\href{http://arxiv.org/abs/1211.6767}{{\tt
  arXiv:1211.6767}}].

\bibitem{Papadodimas:2013jku}
K.~Papadodimas and S.~Raju, {\it {State-Dependent Bulk-Boundary Maps and Black
  Hole Complementarity}},  {\em Phys. Rev.} {\bf D89} (2014), no.~8 086010,
  [\href{http://arxiv.org/abs/1310.6335}{{\tt arXiv:1310.6335}}].

\bibitem{Papadodimas:2015jra}
K.~Papadodimas and S.~Raju, {\it {Remarks on the necessity and implications of
  state-dependence in the black hole interior}},  {\em Phys. Rev.} {\bf D93}
  (2016), no.~8 084049, [\href{http://arxiv.org/abs/1503.0882}{{\tt
  arXiv:1503.0882}}].

\bibitem{Verlinde:2012cy}
E.~Verlinde and H.~Verlinde, {\it {Black Hole Entanglement and Quantum Error
  Correction}},  {\em JHEP} {\bf 10} (2013) 107,
  [\href{http://arxiv.org/abs/1211.6913}{{\tt arXiv:1211.6913}}].

\bibitem{Verlinde:2013uja}
E.~Verlinde and H.~Verlinde, {\it {Passing through the Firewall}},
  \href{http://arxiv.org/abs/1306.0515}{{\tt arXiv:1306.0515}}.

\bibitem{Verlinde:2013qya}
E.~Verlinde and H.~Verlinde, {\it {Behind the Horizon in AdS/CFT}},
  \href{http://arxiv.org/abs/1311.1137}{{\tt arXiv:1311.1137}}.

\bibitem{Maldacena:2013xja}
J.~Maldacena and L.~Susskind, {\it {Cool horizons for entangled black holes}},
  {\em Fortsch. Phys.} {\bf 61} (2013) 781--811,
  [\href{http://arxiv.org/abs/1306.0533}{{\tt arXiv:1306.0533}}].

\bibitem{Hamilton:2006az}
A.~Hamilton, D.~N. Kabat, G.~Lifschytz, and D.~A. Lowe, {\it {Holographic
  representation of local bulk operators}},  {\em Phys. Rev.} {\bf D74} (2006)
  066009, [\href{http://arxiv.org/abs/hep-th/0606141}{{\tt hep-th/0606141}}].

\bibitem{Faulkner:2017vdd}
T.~Faulkner and A.~Lewkowycz, {\it {Bulk locality from modular flow}},  {\em
  JHEP} {\bf 07} (2017) 151, [\href{http://arxiv.org/abs/1704.0546}{{\tt
  arXiv:1704.0546}}].

\bibitem{Almheiri:2017fbd}
A.~Almheiri, T.~Anous, and A.~Lewkowycz, {\it {Inside out: meet the operators
  inside the horizon. On bulk reconstruction behind causal horizons}},  {\em
  JHEP} {\bf 01} (2018) 028, [\href{http://arxiv.org/abs/1707.0662}{{\tt
  arXiv:1707.0662}}].

\bibitem{Gao:2016bin}
P.~Gao, D.~L. Jafferis, and A.~Wall, {\it {Traversable Wormholes via a Double
  Trace Deformation}},  {\em JHEP} {\bf 12} (2017) 151,
  [\href{http://arxiv.org/abs/1608.0568}{{\tt arXiv:1608.0568}}].

\bibitem{Maldacena:2017axo}
J.~Maldacena, D.~Stanford, and Z.~Yang, {\it {Diving into traversable
  wormholes}},  {\em Fortsch. Phys.} {\bf 65} (2017), no.~5 1700034,
  [\href{http://arxiv.org/abs/1704.0533}{{\tt arXiv:1704.0533}}].

\bibitem{Kourkoulou:2017zaj}
I.~Kourkoulou and J.~Maldacena, {\it {Pure states in the SYK model and
  nearly-$AdS_2$ gravity}},  \href{http://arxiv.org/abs/1707.0232}{{\tt
  arXiv:1707.0232}}.

\bibitem{Cardy:2004hm}
J.~L. Cardy, {\it {Boundary conformal field theory}},
  \href{http://arxiv.org/abs/hep-th/0411189}{{\tt hep-th/0411189}}.

\bibitem{Liendo:2012hy}
P.~Liendo, L.~Rastelli, and B.~C. van Rees, {\it {The Bootstrap Program for
  Boundary CFT$_d$}},  {\em JHEP} {\bf 07} (2013) 113,
  [\href{http://arxiv.org/abs/1210.4258}{{\tt arXiv:1210.4258}}].

\bibitem{McAvity:1995zd}
D.~M. McAvity and H.~Osborn, {\it {Conformal field theories near a boundary in
  general dimensions}},  {\em Nucl. Phys.} {\bf B455} (1995) 522--576,
  [\href{http://arxiv.org/abs/cond-mat/9505127}{{\tt cond-mat/9505127}}].

\bibitem{Takayanagi:2011zk}
T.~Takayanagi, {\it {Holographic Dual of BCFT}},  {\em Phys. Rev. Lett.} {\bf
  107} (2011) 101602, [\href{http://arxiv.org/abs/1105.5165}{{\tt
  arXiv:1105.5165}}].

\bibitem{Fujita:2011fp}
M.~Fujita, T.~Takayanagi, and E.~Tonni, {\it {Aspects of AdS/BCFT}},  {\em
  JHEP} {\bf 11} (2011) 043, [\href{http://arxiv.org/abs/1108.5152}{{\tt
  arXiv:1108.5152}}].

\bibitem{DeWolfe:2001pq}
O.~DeWolfe, D.~Z. Freedman, and H.~Ooguri, {\it {Holography and defect
  conformal field theories}},  {\em Phys. Rev.} {\bf D66} (2002) 025009,
  [\href{http://arxiv.org/abs/hep-th/0111135}{{\tt hep-th/0111135}}].

\bibitem{Alishahiha:2011rg}
M.~Alishahiha and R.~Fareghbal, {\it {Boundary CFT from Holography}},  {\em
  Phys. Rev.} {\bf D84} (2011) 106002,
  [\href{http://arxiv.org/abs/1108.5607}{{\tt arXiv:1108.5607}}].

\bibitem{Shenker:2014cwa}
S.~H. Shenker and D.~Stanford, {\it {Stringy effects in scrambling}},  {\em
  JHEP} {\bf 05} (2015) 132, [\href{http://arxiv.org/abs/1412.6087}{{\tt
  arXiv:1412.6087}}].

\bibitem{Heemskerk:2012mn}
I.~Heemskerk, D.~Marolf, J.~Polchinski, and J.~Sully, {\it {Bulk and
  Transhorizon Measurements in AdS/CFT}},  {\em JHEP} {\bf 10} (2012) 165,
  [\href{http://arxiv.org/abs/1201.3664}{{\tt arXiv:1201.3664}}].

\bibitem{HOOFT1987}
G.~Hooft, {\it Graviton dominance in ultra-high-energy scattering},  {\em
  Physics Letters B} {\bf 198} (1987), no.~1 61 -- 63.

\end{thebibliography}\endgroup
\end{document}